\documentclass[preprint1]{aastex}

\usepackage{graphicx, natbib, longtable, lscape, amssymb}
\usepackage[english]{babel}
\usepackage{epsfig}
\usepackage{amssymb,amsmath}
\usepackage{lscape}
\usepackage{rotating}

\newcommand{\gppr}{\stackrel{>}{\scriptstyle \sim}}
\newcommand{\gappr}{\raisebox{-0.4ex}{$\gppr$}}
\newcommand{\lppr}{\stackrel{<}{\scriptstyle \sim}}
\newcommand{\lappr}{\raisebox{-0.4ex}{$\lppr$}}

\newcommand{\lto}{$\lambda_{\mathrm{turn−off}}$}
\newcommand{\afex}{$\alpha_{\mathrm{excess}}$}
\newcommand{\Msun}{M$_{\odot}$}
\newcommand{\Lsun}{L$_{\odot}$}
\newcommand{\Md}{M$_{\mathrm{DISK}}$}

\newcommand{\Mjup}{M$_{\mathrm{JUP}}$}
\newcommand{\Ld}{L$_{\mathrm{DISK}}$}
\newcommand{\Ls}{L$_{*}$}

\slugcomment{To appear in ApJ}
\shorttitle{The Nature of Transition Circumstellar Disks II}
\shortauthors{Romero, G.A. et al.}

\begin{document}

\title{The Nature of Transition Circumstellar Disks II.\\Southern Molecular Clouds{\LARGE{$^{\star}$}}}

\author{Gisela A. Romero\altaffilmark{1,2,3},
Matthias R. Schreiber\altaffilmark{1},
Lucas A. Cieza\altaffilmark{4},
Alberto Rebassa-Mansergas\altaffilmark{1},
Bruno Mer\'in\altaffilmark{5},
Anal\'ia V. Smith Castelli\altaffilmark{3,6},
Lori E. Allen\altaffilmark{7}, and 
Nidia Morrell\altaffilmark{8}
}

\altaffiltext{}{$\star$ Based in part on observations made with ESO telescopes at Paranal and APEX Observatories, under ESO programs 083.C-0459(A), 085.C-0571(D), 083.F-0162(A). This paper includes data gathered with the 6.5-m Magellan Telescopes located at Las Campanas Observatory.}
\altaffiltext{1}{Departamento de F\'isica y Astronom\'ia, Universidad de Valpara\'{\i}so, Valpara\'{\i}so, Chile}
\altaffiltext{2}{C\'atedra de Medio Interestelar, Facultad de Ciencias Astron\'omicas y Geof\'{\i}sicas, Universidad Nacional de La Plata, La Plata, Argentina}
\altaffiltext{3}{Consejo Nacional de Investigaciones Cient\'{i}ficas y
  T\'{e}cnicas (CONICET), Argentina}
\altaffiltext{4}{Institute for Astronomy, University of Hawaii at Manoa,
  Honolulu, HI 96822, \em{Sagan} Fellow}
\altaffiltext{5}{Herschel Science Centre, ESAC (ESA), P.O. Box 78, 28691 Villanueva de la Ca\~nada, Madrid, Spain}
\altaffiltext{6}{Instituto de Astrof\'isica de La Plata (CCT La Plata-CONICET-UNLP), Paseo del Bosque, B1900 FWA La Plata, Argentina}
\altaffiltext{7}{Department of Astronomy, University of Arizona, 933 N Cherry Ave., Tucson, AZ 85721-0065, USA}
\altaffiltext{8}{Las Campanas Observatory, Carnegie Observatories, Casilla 601, La Serena, Chile}

\begin{abstract}
Transition disk objects are pre-main-sequence stars with
little or no near-IR excess and significant far-IR excess,
implying inner opacity holes in their disks. Here we present a multifrequency
study of transition disk candidates located in
Lupus I, III, IV, V, VI, Corona Australis, and Scorpius. Complementing the information provided by
{\em{Spitzer}} with Adaptive Optics (AO) imaging (NaCo, VLT), submillimeter
photometry (APEX), and echelle spectroscopy (Magellan, Du Pont
Telescopes), we estimate the multiplicity, disk mass, and accretion rate for
each object in our sample in order to identify the mechanism potentially
responsible for its inner hole. We find that our transition disks show a rich
diversity in their SED morphology, have disk masses ranging 
from $\lappr$ 1 to 10
\Mjup~and accretion rates ranging from $\lappr$ $10^{-11}$ to
$10^{-7.7}$ M$_{\odot}$ yr$^{-1}$. Of the 17 bona fide transition disks in our sample, 3, 9, 3, and 2 objects are consistent with giant planet formation, grain growth, 
photoevaporation, and debris disks, respectively. 
Two disks could be circumbinary, 
which offers tidal truncation as an alternative origin of the inner hole.
We find the same heterogeneity of the transition disk
population in Lupus III, IV, and Corona Australis as in our previous
analysis of transition disks in Ophiuchus while all transition disk candidates
selected in Lupus V, VI turned out to be contaminating background AGB stars.
All transition disks classified as photoevaporating disks have small disk 
masses, which indicates that
photoevaporation must be less efficient than predicted by most recent models. 
The three systems that are excellent candidates for harboring giant planets
potentially represent invaluable laboratories to study planet 
formation with the Atacama Large Millimeter/Submillimeter Array..

\end{abstract}

\section{Introduction}\label{intro}

Low-mass pre-main-sequence (PMS) stars are generally separated in two different classes, 
accreting classical T\,Tauri stars (CTTSs) with broad
H$\alpha$ emission lines, 
blue continuum and near infrared excess 
and non-accreting weak-line T\,Tauri Stars (WTTSs) with 
narrow symmetric H$\alpha$ emission lines
\citep[e.g.][]{bertout84-1}. 
While CTTSs typically show large excess 
emission from the near-infrared to 
the millimeter, WTTSs often have no
infrared (IR) excess at all. 
Only a relatively small fraction of T\,Tauri stars are observed
in an intermediate transition state with little or no near-IR 
excess and significant far-IR excess. 
This clearly indicates that once the inner disk starts to dissipate, the entire
disk disappears very rapidly \citep{wolk+walter96-1,andrews+williams05-1,ciezaetal07-1}.
The missing near-IR excess combined with the clear 
presence of an outer disk is the
defining characteristic of transition disks. However, a precise and generally accepted definition of what 
constitutes a transition disk object does not yet exist. 
The most conservative definition of transition disks, often labeled
{\it classical transition disks}, consists of objects with no detectable 
near-IR excess, 
steeply rising slopes in the mid-IR, and large far-IR excesses 
\citep[e.g.][]{muzzerolleetal06-1,Sicilia-Aguilaretal06-1,muzzerolleetal10-1}.
Being less restrictive, objects with small, but still detectable,
near-IR excesses \citep[e.g.][]{brownetal07-1,merinetal10-1} can be
included, until considering objects with decrement relative to the Taurus
median Spectral Energy Distribution (SED) at any or all wavelengths
\citep[e.g.][]{najitaetal07-1,ciezaetal10-1}. Throughout this paper we follow
the latter and broader definition.
However, one has to be aware that this broad definition still is
mostly sensitive to inner opacity holes but may overlook 
pre-transitional disks with a gap separating an optically thick inner 
disk from an optically thick outer disk. Such systems have been 
identified from Spitzer IRS spectra 
\citep{espaillatetal07-1}, but can be missed by photometric 
selection alone.

The {\em{Spitzer}} Space Telescope generated a huge database 
containing IR observations of PMS stars
in star-forming regions. Most importantly, 
{\em{Spitzer}} products such as the catalogs of the {\em{Cores
to Disks (c2d)}}\footnote{http://irsa.ipac.caltech.edu/data/SPITZER/C2D/doc/c2d$\_$del$\_$document.pdf}
 and  {\em{Gould  Belt Spitzer (GB)}} Legacy 
projects \citep{spezzietal11-1,petersonetal11-1}
provide SEDs from 3.6 to 24 $\mu$m for large numbers of 
PMS stars. One of the most interesting results concerning 
transition disk studies with
{\em{Spitzer}} has been the great diversity of SED morphologies \citep[see][for a review]{williams+cieza11-1}.
The widespread of IR SED morphologies found in transition disk objects
cannot be adapted to the  
classical taxonomy to describe young stellar objects (YSOs) such as the Class\,
 I, II, III definitions from \citet{lada87-1}.   
\cite{ciezaetal07-1} quantified the richness of SED morphologies in terms of
two parameters based on the SED shapes considering the longest 
wavelength at which the observed flux is dominated by
the stellar photosphere, \lto, and the slope of the
infrared excess, \afex, computed from~\lto~to 24 $\mu$m. 

Studying the diverse population of transition disks is key 
for understanding circumstellar disk evolution as much of the diversity of
their SED morphologies is likely  
to arise from different physical processes dominating the disk's evolution.
Evolutionary processes that may play an important role include
viscous accretion 
\citep{hartmannetal98-1},
photoevaporation \citep{alexanderetal06-1}, the magneto-rotational instability ({\em{MRI}})
\citep{chiang+murray-clay07-1}, grain growth and dust settling
\citep{dominik+dullemond08-1}, planet formation \citep{lissauer93-1,boss00-1}
and dynamical interactions between the disk and stellar or substellar
companions \citep{artymowicz+lubow94-1}. 

As discussed by \citet{najitaetal07-1,cieza08-1,alexander08-1}, one can
distinguish between some of these processes if certain observational constraints, 
in addition to the SEDs, are available. 
To this end, we are performing an extensive 
ground-based observing program to obtain estimates for the 
disk masses (from submillimeter photometry), accretion rates 
(from the velocity profiles of the H$\alpha$ line), and multiplicity
information (from AO observations) of {\em{Spitzer}}-selected disks in several nearby
star-forming regions. 
Our recently completed 
study of Ophiuchus objects \citep[][hereafter Paper~I]{ciezaetal10-1} 
confirms that transition disks are indeed a very 
heterogeneous group of objects with a wide range of SED morphologies, disk
masses ($<$ 0.5 to 40 \Mjup) and accretion rates \hbox{($<$ 10$^{-11}$ to
10$^{-7}$ M$_{\odot}$ yr$^{-1}$)}. Since the properties of the transition
disks in our sample point towards 
different processes driving the evolution of each disk, 
we have been able to identify strong candidates for
the following disk categories: (giant) planet-forming disks, circumbinary
disks, grain-growth dominated disks, 
photoevaporating disks, and debris disks.

We here follow the same approach as in Paper I in performing multiwavelength
observations to derive estimates on disk masses, accretion rates, and
multiplicity. We present submillimeter wavelength photometry (from APEX),  
high-resolution optical spectroscopy (from the Clay, and Du Pont telescopes), 
and Adaptive Optics near-IR imaging  (from the VLT) for {\em{Spitzer}}-selected transition circumstellar disks located  
in the following star forming regions: a) Lupus: I, III, IV, V, VI, b) Corona Australis (CrA), and c) Scorpius (Scp).

\section{Transition disks in Southern star-forming regions}\label{selection}

The Lupus clouds constitute one of the main southern nearby low-mass star-forming regions containing the following sub-clouds at 
slightly different distances: Lupus I, IV, V, VI at 150 $\pm$ 20 pc and Lupus III at 200 $\pm$ 20 pc  
\citep{comeron08-1}. The clouds are situated in the Lupus-Scorpius-Centaurus OB association spanning over 20 deg in the sky.
Their population is dominated by mid M-type PMS stars, 
but some very late M stars or substellar
objects have been found as well 
thanks to {\em{Spitzer}} capabilities \citep[see][for a
review]{comeron08-1}.    
In general, the ages of the Lupus clouds are estimated to be 
$\approx\,1.5-4$ Myr, \citep{hughesetal94-1,comeronetal03-1}. 
However, a comprehensive analysis using {\em{Spitzer}} IRAC and
MIPS observations in combination with near-IR (2MASS) data has been
performed for Lupus I, III, and IV  by the {\em{c2d}} Legacy Project \citep{merinetal08-1} 
and Lupus\, V, VI  by the {\em{Gould Belt}} Legacy Project
\citep{spezzietal11-1} and revealed a significant difference between the
sub-clouds. 
While Lupus I, III, and IV are dominated by Class II YSOs, 
Lupus\, V, VI mostly contain Class III objects. This 
has been interpreted as
a consequence of Lupus\, V, VI being a few Myrs older than 
Lupus I, III, and IV by \citet{spezzietal11-1}.
In any case, the Lupus star-forming regions represent 
an excellent test-bed for theories of circumstellar disk evolution
as their stellar members should span all evolutionary stages.  

The Scorpius clouds \citep{nozawaetal91-1,vilasboasetal00-1} lie on the 
edge of the  Lupus-Scorpius-Centaurus OB association, just north of the well
studied Ophiuchus molecular cloud, but it is highly fragmentary and
presents much lower levels of star-formation. In fact, the {\em{Gould Belt}} project only finds 10 YSOs 
candidates in
the 2.1 sq deg. mapped by IRAC and MIPS (Hatchell et al., in preparation).  
The age of Scp is estimated to be $\sim\,5$~Myr
\citep{preibischetal02-1}.

The CrA star-forming region, also mapped by the  {\em{Gould Belt}} project
\citep{petersonetal11-1}, contains an embedded association known as the
Coronet, a relatively isolated cluster containing 
HAeBe stars and T\,Tauri stars \citep{chenetal97-1}. 
It is situated at a distance of 150 $\pm$ 20 pc 
out of the Galactic plane, at the edge of the {\em{Gould Belt}}
\citep[see][and references therein]{Sicilia-Aguilaretal08-1}. 
With an age of $\approx$ 1 Myr, the Coronet is younger than the Lupus clouds
and has been claimed to host an intriguingly  
high fraction of classical transition disks of $\approx~50\%$
 \citep{Sicilia-Aguilaretal08-1}. 
However, \citet{ercolanoetal09-1}
 convincingly shows that the dust emission 
in T\,Tauri stars of spectral type M is very small short ward of 
$6\mu$m which might mimic an inner hole, the defining feature of
typical transition disk systems. Based on this finding,
\citet{ercolanoetal09-1} 
estimate a much smaller fraction of transition disks 
in Coronet, of $\sim15\%$.

\subsection{Target selection}

We have systematically searched the catalogs of the {\em{c2d}} and
{\em{Gould Belt}} Legacy Projects\footnote{the former is available at   http://irsa.ipac.caltech.edu/data/SPITZER/C2D/}
applying the broad transition disk definition described in detail in Paper I
to the Lupus I, III, IV, V, VI, Scp, and CrA clouds. 
In brief, we select systems that fulfill the following criteria.

\begin{enumerate} 

\item Have Spitzer colors [3.6]-[4.5] $<$ 0.25, which excludes
``fulls disks", i.e., optically thick disks extending inward to the dust
sublimation radius except in cases with significant
dust settling in inner disks around M stars \citep{ercolanoetal09-1}.

\item Have Spitzer colors [3.6]-[24] $>$ 1.5, to ensure
that all targets have very significant excesses ($>$ 5-10 $\sigma$), 
unambiguously indicating the presence of circumstellar material.

\item Have S/N $\ge$ 7 in 2MASS, IRAC, MIPS
(24 $\mu$m) bands to only 
include targets with reliable photometry. 

\item Have K$_{s}$ $<$ 11 mag, driven by the sensitivity of our 
near-IR Adaptive Optics observations and 
to avoid extragalactic contamination. 

\item Are brighter than R = 18 mag according to the USNO-B1
  \citep{monetetal03-1}, driven by the sensitivity of our optical spectroscopy
  observations. Compared to  the c2d sample
  discussed in \citet{merinetal08-1} our sample might be slightly biased
  against very low mass stars and deeply embedded objects because of this
 brightness limit.

\end{enumerate} 
These selection criteria result in a primary target list of 60 objects that we
did follow-up using different observational facilities to characterize our
transition disk candidate sample.

\section{Observations}\label{obs}

We performed multiwavelength (optical, infrared, and submillimeter) 
observations of our targets with the aim to identify 
which physical process is primarily responsible for their transition disk 
nature. 
High resolution optical spectra 
can be used to estimate spectral types and accretion rates 
from the velocity dispersion of the H$\alpha$.
Near-IR images allow to identify multiple star systems down to projected separations of 
0.06-007$''$, corresponding to 
$8-14$~AU at distances of $130-200$~pc. 
From single dish submillimeter observations we inferred  disk masses. 
In the following Section, we describe in detail the observations 
performed and the data reduction. 

\subsection{Optical Spectroscopy}

We obtained high resolution (R $>$ 20,000) spectra for our 
entire sample using 2 different telescopes: Magellan/Clay and Du Pont located 
at Las Campanas Observatory in Chile.

\subsubsection{Clay--Mike Observations}

We observed 49 of our 60 targets with the Magellan Inamori Kyocera
Echelle (MIKE) spectrograph on the 6.5-m Clay telescope.  The
observations were performed on 2009 April 27--28 and 2010 June
11--13. Since the CCD of MIKE's red arm has a pixel scale of
  0.13$''$/pixel, we binned the detector by a factor of 3 in the
  dispersion direction and a factor of 2 in the spatial direction,
  thus reducing the readout time and readout noise. We used an 1$''$
slit width. The resulting spectra covered 4900--5000\AA\, at a
  resolving power of 22,000. This corresponds to a resolution
  of $\sim$0.3 \AA\, at the location of the H$\alpha$ line, and to a
velocity dispersion of $\sim$14~km s$^{-1}$.

For each object, we obtained a set of 3 or 4 spectra, with exposure
times ranging from 3 to 10 minutes each, depending on the brightness
of the targets. The data analysis was carried out with
IRAF\footnote{Image Reduction and Analysis Facility, distributed by
  NOAO, operated by AURA, Inc., under agreement with NSF} software.
After bias subtraction and flat-field corrections with Milky Flats,
the spectra were reduced using the standard IRAF package
IMRED:ECHELLE.

\subsubsection{Du Pont--Echelle  Observations}

The remaining 11 targets were observed with the Echelle Spectrograph
on the 2.5-m Ir\'en\'ee du Pont telescope. The observations were
performed in 2009 May 14--16, and we used an 1$''$ slit
  width. The CCD's scale is 0.26$''$/pixel, and we consequently
  applied a 2$\times$2 binning. The wavelength coverage of the
  obtained spectra ranged between 4000 and 9000 \AA\, at a resolving
  power of 32,000 in the red arm. This corresponds to a resolution of
  $\sim$0.2\AA\, and a velocity dispersion of $\sim$9.4 km s$^{-1}$
  in the vicinity of H$\alpha$.

% We used a 1$''$ slit to obtain spectra
%between 4000 and 9000 \AA \ with a resolution of 45,000, corresponding
%to 0.14 \AA~in the red. However, since the Spectrograph's CCD has a
%pixel scale of $\sim$0.1 \AA/pixel, the true two-pixel resolution
%corresponds to $\sim$32,000, or $\sim$9.4 km s$^{-1}$ in the red.

For each object we obtained a set of 3 to 4 spectra with exposure
times ranging from 10 to 15 minutes each, depending on the brightness
of the target. The data analysis was carried out with IRAF. After bias
subtraction and flat-field corrections with Milky Flats, the spectra
were reduced using the standard package IMRED:ECHELLE.

\subsection{Adaptive Optics Imaging}\label{s:NaCo}

High spatial resolution near-IR observations of our 60 targets were
obtained  
with NaCo (the Nasmyth Adaptive Optics Systems (NAOS) and the Near-IR Imager and Spectrograph (CONICA) camera at  
the 8.2-m telescope Yepun), which is part of the European Southern Observatory's 
(ESO) Very Large Telescope (VLT) in Cerro Paranal, Chile.  
The data were acquired in service mode during 
the ESO's observing period 083 (2009 April 1 -- September 30).  

To take advantage of the near-IR brightness of our targets, we used 
the infrared wavefront sensor and the N90C10 dichroic to direct 90$\%$ of the 
near-IR light to the adaptive optics systems 
and 10$\%$ of the light to the science 
camera. We used the S13 camera (13.3 mas/pixel and 14$\times$14$''$ field of 
view) and the Double RdRstRd readout mode. The observations were 
performed through the K$_s$ and J-band filters at 5 dithered positions per 
filter. The total exposure times ranged from 1 to 50 s for  the K$_s$-band
observations and from 2 to 200\,s 
for the J-band observations, depending on the brightness
of the target.  
The data were reduced  using the Jitter software, which is part of ESO's data
reduction package  
Eclipse\footnote{\tt http://www.eso.org/projects/aot/eclipse/ }.

\subsection{Submillimeter Wavelength Photometry}

As discussed in the following section,  our spectroscopic observations 
showed that our 
initial sample of 60 transition disk candidates was highly contaminated by asymptotic giant branch (AGB)
stars. The 17 bona fide PMS stars were observed with the 
Atacama Pathfinder Experiment (APEX)
\footnote{This publication is based on data acquired with 
APEX which is a collaboration between the Max-Planck-Institut fur 
Radioastronomie, the European Southern Observatory, and the Onsala Space 
Observatory.}, the 12-m radio telescope located in Llano de Chajnantor in
Chile. The observations were performed during 
period 083 (083.F-0162A-2009, 9.2 hrs) and period 085 
(E-085.C-0571D-2010, 30.9 hrs).
We used the APEX-LABOCA camera \citep{siringoetal09-1} at 870 $\mu$m (345 GHz)
in service mode aiming for detections of the dust continuum emission. 
The nominal LABOCA beam is full width at
half-maximum $18.6\pm1.0$'' and the pointing uncertainty 
is $4$''. To obtain the lowest possible flux limit, 
the most sensitive part of the array was centered on each source. 
The observations were reduced using the Bolometer array
data Analysis package BoA$\footnote{http://www.apex-telescope.org/bolometer/laboca/boa/}$. 

For both observing runs, Skydips were performed hourly and combined with 
radiometer readings to obtain accurate opacity estimates. The absolute flux
calibration follows the method outlined by \cite{siringoetal09-1} and is
expected to be accurate to within 10$\%$. The absolute flux scale
pointing calibrators were determined through observations of either
IRAS16342-38 or G34.3 while planets were used to focus the telescope. 
The telescope pointing was
checked regularly with scans on nearby bright sources and was found to 
be stable within 3'' (rms). 

The period 083 observations were performed using compact mapping mode with raster spiral
patterns. The weather conditions were excellent with precipitable water vapor
levels below $0.5$~mm. Eight sources 
(objects \# 1, 2, 5, 9, 12, 15, 16, 17)
were observed. On-source integrations of 64 minutes were performed to achieve an
rms $\sim$ 7 mJy/beam. 
The brightest object of the whole sample (\# 2) was the only source detected 
at submillimeter wavelengths in period 083. 
During the longer period 085 observing run, the beam switching mode using the
wobbling secondary and mapping mode were employed. 
During this period, the remaining nine sources were observed and 
object \#12 was re-observed with higher sensitivity.
The weather conditions were favorable with precipitable water vapour levels
below 1.2~mm. The wobbler observations of each target consist of a set of 
two loops of 10 scans per target, reaching a total on-source observing time of 
48 minutes. An average rms $\sim$ 4 mJy/beam was obtained. In the case of a 
signal detection
on-source position, we took a few maps in order to check for emission 
contamination from the off-position. In all cases, the contamination 
was discarded and we confirmed the detection of six sources
(\# 3, 7, 8, 10, 11, and 12).

\section{Results}

\subsection{AGB Contamination} \label{contamination}

AGB stars are surrounded by shells of dust and thus 
have small, but detectable, IR excesses.   
The {\em{Spitzer}}-selected YSO samples 
from {\em{c2d}} and {\em{Gould Belt}} catalogs
are therefore contaminated by 
AGB stars. 
Using high resolution optical spectra, we discovered that 43 objects
of our candidates are AGB stars, while the remaining 17 targets are
spectroscopically confirmed T\,Tauri stars. 
We separated contaminating AGB stars from genuine transition disk T\,Tauri
stars in the same way as in Paper I, i.e. based on the presence/absence of 
emission lines associated with chromospheric activity and/or accretion
and the presence of  the Li  6707 \AA~absorption line indicating stellar youth. 
The coordinates, {\em{Spitzer}} names, the USNO-B1 R-band magnitude, 
and the near 
to mid-IR fluxes of the AGB stars contaminating our sample of transition disks 
are compiled in Table \ref{t:giants}. 

As shown in Table \ref{t:cont} the 
fractional contamination due to AGB stars of our color selected transition disk
candidates  differs significantly between the different clouds. 
The number of transition disk candidates is far too small 
in the case of Lupus I and Scp to draw any conclusions. 
Our Lupus III, IV, and CrA samples are contaminated by a fraction 
of AGB stars that
is more or less consistent with the contamination in Ophiuchus (see
Paper\,I, section 4.1.2). 

The small number of transition disks in CrA 
seems to be in contradiction with the larger
sample identified by \citet{Sicilia-Aguilaretal08-1}. 
However, our selection criteria contain relatively strong brightness 
constraints (in particular $K<11$)
due to the design of our follow-up program which excludes most of
the systems listed by them. 
In addition, as mentioned in the introduction, a large fraction of 
the transition disk candidates of
\citet{Sicilia-Aguilaretal08-1} might be classical M-dwarf T\,Tauri stars 
with intrinsically little near IR excess due to the small color contrast
between the disk and the stellar photosphere \citep{ercolanoetal09-1}.

Apparently, Lupus\,V and VI are dramatically more
contaminated than Lupus III, IV, i.e. all the color-selected transition disk
candidates are in fact AGB stars. 
This high percentage of contamination is
perhaps related to the position in the Galaxy (see Table \ref{t:cont}). 
The Lupus complex occupies 334 $<$ l $<$ 352, +5
$<$ b $<$ + 25, i.e. observing Lupus\,V and VI we are looking towards the 
galactic center closer to the plane.  In contrast, CrA and Ophiuchus (Paper
I) are located at higher Galactic latitudes.
In any case, the absence of any spectroscopically 
confirmed transition disk in Lupus\,V, VI
puts doubts on the 
finding of \citet[][see their section 5.1]{spezzietal11-1} 
that the high fraction of Class\,III Lada systems 
can {\em{not}} be explained by contamination. So far all Class\,III YSO
candidates from these clouds that have been followed up spectroscopically 
are clearly contaminating background giants. 
Our sample of transition disks in Lupus\, V, VI 
shares 30 Class\,III objects and one Class\,II object with the sample
investigated by \citet{spezzietal11-1}. All these 31 objects 
turned out to be AGB stars which means that at 
least $\sim50\%$ and potentially much
more of the Class\,III objects from \citet{spezzietal11-1} are not YSOs 
but giant stars. This result also questions the conclusion of 
\citet{spezzietal11-1} that Lupus V, VI are significantly older than 
Lupus I, III.

\subsection{Color selection of AGB star candidates}

\begin{figure}[t]
\begin{center}
\includegraphics[width=5.0in]{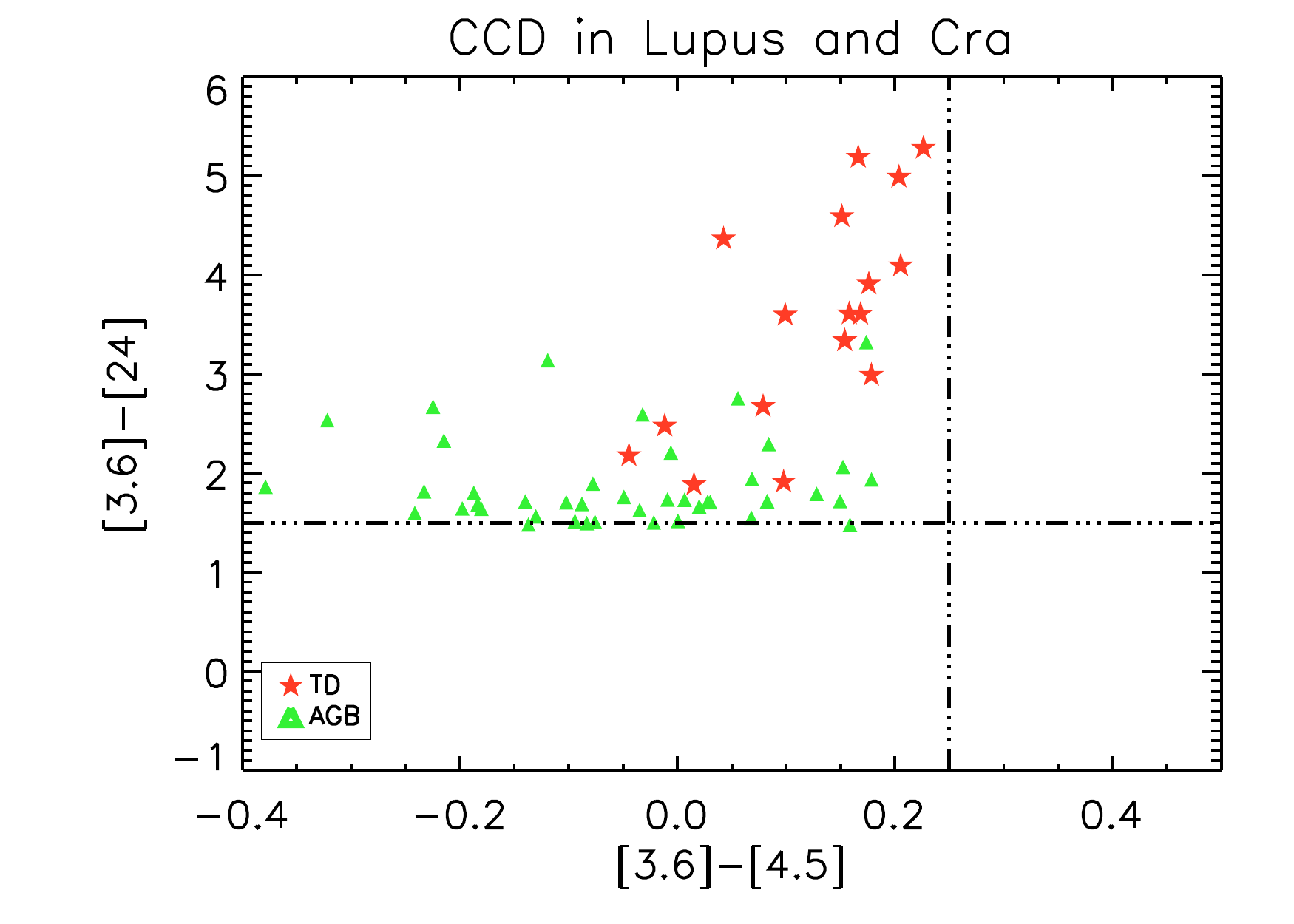}
\end{center}
\caption{
Spitzer color-color diagram for the southern sample. Contaminating AGB stars 
greatly dominate the sample with \hbox{[3.6] – [24] $<$ 1.8}.
}
\label{f:ccont}
\end{figure}

\begin{figure}[t]
\includegraphics[width=5.0in]{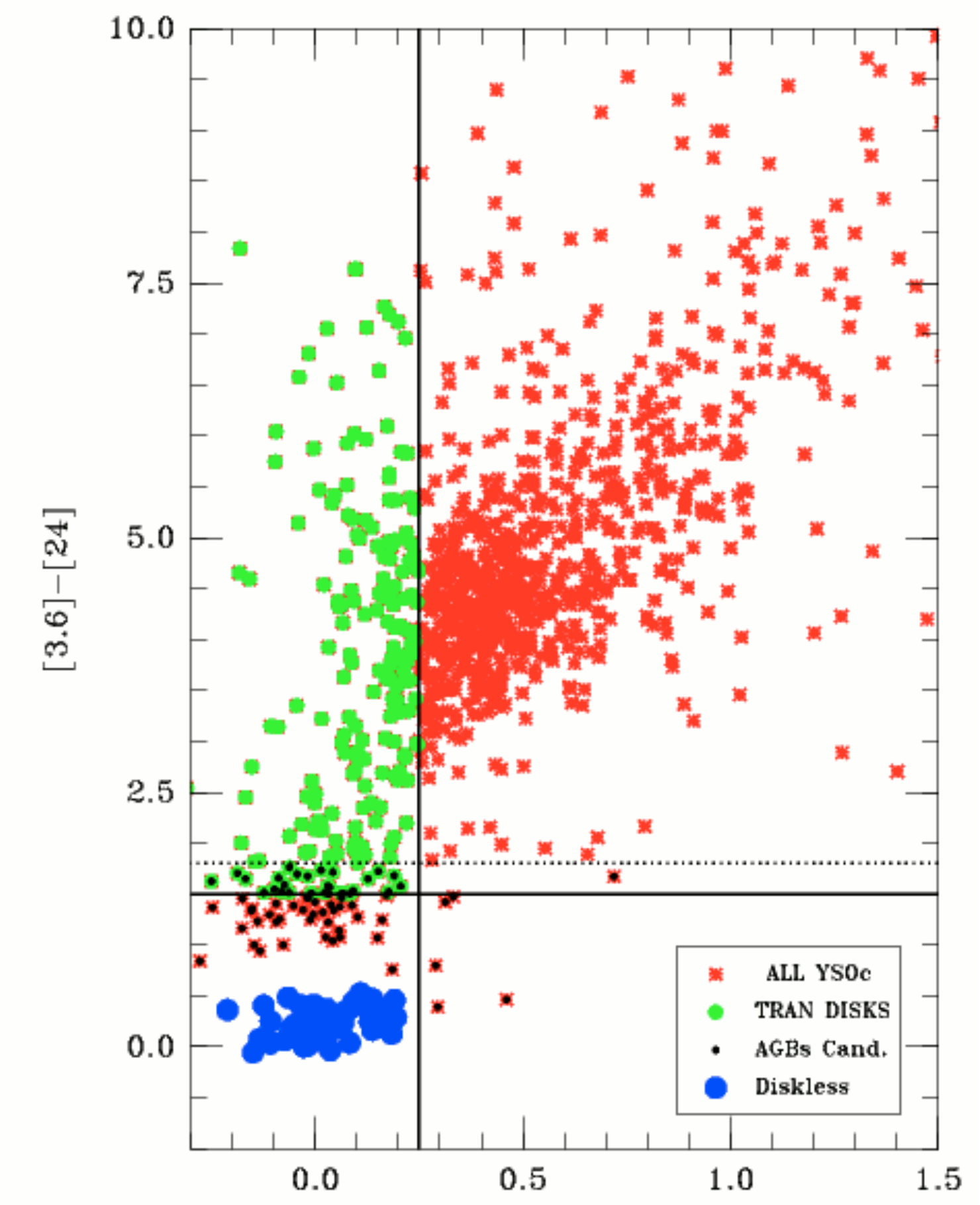}
\caption{\label{f:c2d}
Spitzer {\em{c2d}} systems classified into YSO candidates, AGB candidates,
  and transition disk candidates according to simple color cuts based on the
  results of our spectroscopic follow-up program (see text for more details).  
}
\end{figure}

\begin{figure}[t]
\begin{center}
\includegraphics[width=5.0in]{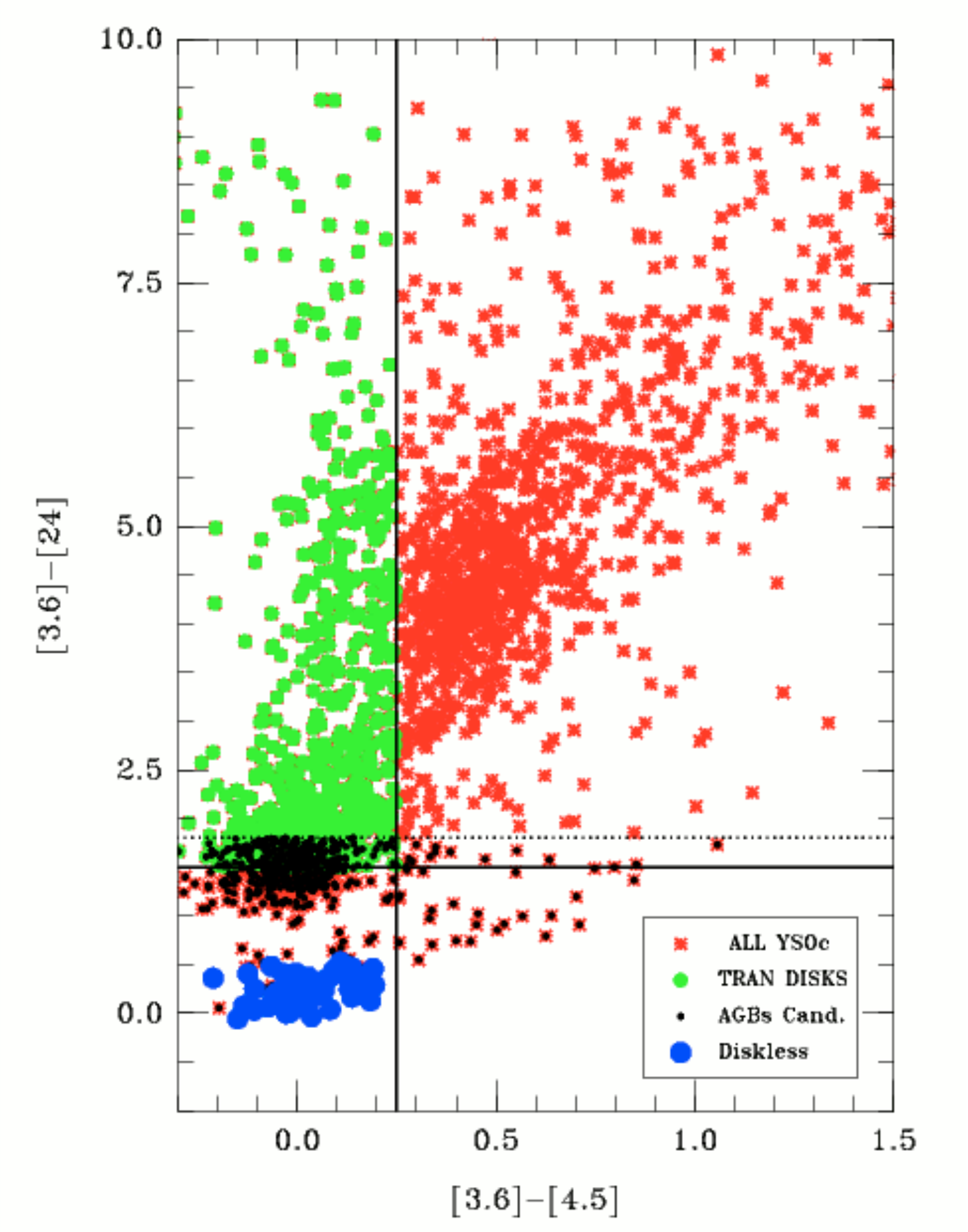}
\end{center}
\caption{\label{f:gb}
Same as Fig.\,\ref{f:c2d} but for the {\em{GB}} catalog. }
\end{figure}

The generally large fraction of giant stars 
in our sample of transition disk
candidates allows to investigate
possible refinements of our color selection algorithm. 
Figure \ref{f:ccont} shows the color-color diagram of the 60 selected 
southern transition disk targets of this paper. 
All transition disk candidates 
in our sample with $[3.6]-[24] < 1.8$ 
turned out to be giant stars.
This agrees quite well with the results obtained for the
Ophiuchus sample where 4/6 transition disk candidates with 
[3.6]-[24] $<$ 1.8 had to be classified as giant stars 
(see Paper\,I, Fig.\,1).  
Consequently, one may derive an estimate of 
the contamination of YSO catalogs due to background 
giant stars by applying this simple color cut. 
Figure\,\ref{f:c2d} and \ref{f:gb} 
show the transition disk candidates and AGB candidates 
for both the {\em{c2d}} and {\em{GB}} catalogs including 
all star forming regions. 
Note, that we here 
apply color selection criteria only, i.e. the 
requirements of $R < 18$ mag  and $K < 11$ mag 
that have been used to define the transition disk candidate 
sample for our multiwavelength follow-up program 
are not incorporated. Instead, here we are interested in  
estimating the fraction of YSO candidates in a given
cloud that are likely to be AGB stars based on their very 
low $24\mu$m excess.  

The resulting rough estimates of giant star 
contamination are given in Table~\ref{giants-all}
separated by catalog 
and cloud. According to these estimates, the AGB contamination 
is expected to vary significantly ranging from 
$\sim1-85$ per cent. This shows that AGB contamination 
can have an important impact on 
studies that are based on the pure numbers of YSOs as provided by the
{\em{c2d}} and {\em{GB}} catalogs. For example, 
star formation rates as determined e.g. in \citet{heidermanetal10-1}
might become significantly smaller if AGB contamination is taken 
into account.

Apparently, applying the new more restrictive color selection 
could also significantly increase the success-rate 
of identifying YSOs directly from color selection criteria
and future follow-up studies may take this into account. 

\subsection{Characterizing southern transition disks}

The {\em{Spitzer}} and alternative 
names, 2MASS and {\em{Spitzer}} fluxes, and  the USNO-B1 R-band 
magnitudes and the relevant information derived 
from our follow-up observations for the remaining 17 
bona fide transition disk candidates are listed in 
Tables \ref{t:mags} and \ref{t:obs}. In what follows, we use 
the data discussed in Section~\ref{obs} to characterize our sample 
of transition disks.

\subsubsection{Spectral types}\label{s:steprop}

In oreder to determine the spectral types of the transition disks in our
sample we use the equations by
\citet{cruz+reid02-1} that empirically relate the spectral type with the
strength of the TiO5 molecular band. 
The uncertainty of this method is estimated to be $\sim$0.5 subclasses. 
For most of our transition disk objects, estimates of the spectral types 
have been provided previously
\citep{hughesetal93-1,hughesetal94-1,krautteretal97-1,walteretal97-1,
Sicilia-Aguilaretal08-1}. 
The spectral types obtained by us and those 
given in the literature are listed in Table \ref{t:obs}
and we find good agreement.
All but one system (\# 2) have been classified as M-dwarfs.  
For target \# 2,  we 
adopt the spectral type K0 given by \citet{hughesetal93-1}.

\subsubsection{Multiplicity}

Binarity can play an important role in the context of transition disks 
as the presence of a close stellar companion may cause the inner hole, i.e.  
some of the transition disks in our sample might actually be nothing else but 
circumbinary disks. Some systems in our sample have been previously 
identified as wide binaries. \citet{merinetal08-1} carried out an optical 
survey of the Lupus I, III, IV regions using bands Rc, Ic, ZwI of the 
Wide-Field Imager (WFI) attached to the ESO 2.2-m telescope at La Silla 
Observatory. Visual inspection of the images 
revealed that objects 
\# 1, \# 3, \# 4, and \# 9 are wide binary systems
with companions at projected separations 
of 420, 1000, 600, and 560 AU considering the distance of the Lupus 
clouds.

\begin{figure}[t]
\begin{center}
\includegraphics[width=5in]{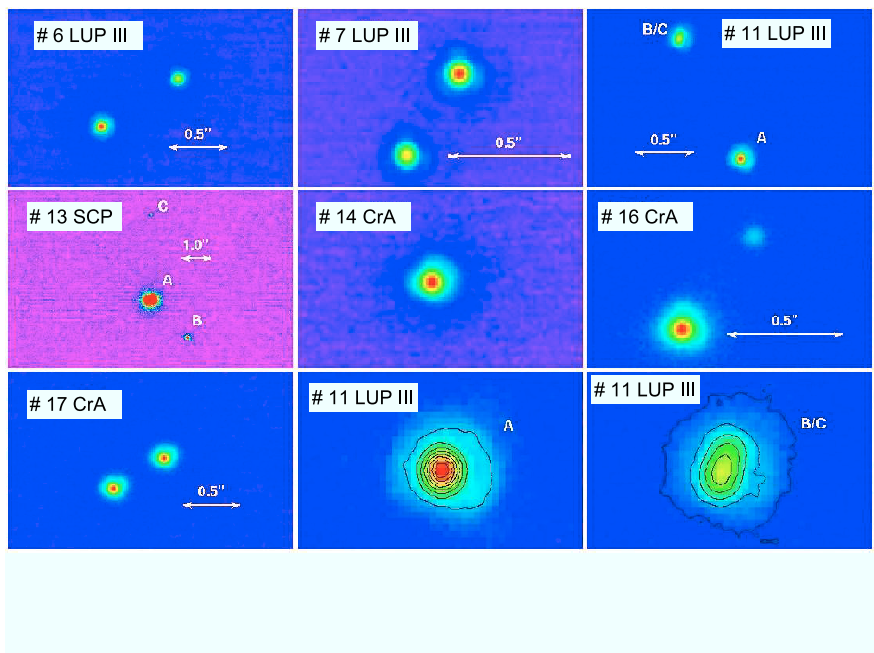}
\end{center}
\caption{The K$_{s}$-band images of the six multiple systems that have been 
detected with our VLT-AO observations 
(objects \# 6, 7, 11, 13, 16, and 17) and of object \# 14, which has been identified as a close binary with
a 0.13$''$ separation from speckle observations in 2001 
\citep{kohleretal08-1}. The putative companion 
in object \# 14 remains unresolved by our 2009 observations. 
Targets \# 11 and 13 are triple systems.  In the former case, the tighter 
components are not fully resolved, but 
their presence can be inferred from the highly elongated 
image (lower right panel).
}
\label{f:mult}
\end{figure}

We have newly identified six multiple systems  
by visual inspection of the NaCo images, 
objects \# 6, \# 7, \# 11, \# 13, \# 16, and \# 17 (see Figure \ref{f:mult}). 
The projected separations are
0.7$''$, 0.4$''$, 1.15$''$, 1.8$''$, 0.5$''$, 0.5$''$ corresponding to 140, 80, 230, 234, 75, 75 AU at the corresponding distances.
Object \# 13 is a triple system, i.e. a binary with an additional 
faint companion at 3$''$ (390 AU). 

For each binary system, we searched for additional 
tight companions by
comparing each other's point-spread functions (PSFs). 
The PSF pairs were virtually identical in all cases, except for
target \# 11. The south-west component of this target has a perfectly 
round PSF, while the
south-east component, $\sim$1.5$''$ away, is clearly elongated
 (see Figure \ref{f:mult}).
  Since variations in the PSF shape
are not expected within such small angular distances and this behavior is seen
in both the J- and K$_{s}$-band images, we conclude 
that target \# 11 is likely to be a triple system.

We have also searched in the literature for additional companions in our sample
that our VLT observations could have missed. In addition to  multiple systems
discussed above, we find that \# 14 has been reported 
by \citet{kohleretal08-1} as a  binary system with a projected
separation of 0.13$''$ (corresponding to 20 AU) and flux ratio of 0.7 using
speckle interferometry at the New Technology Telescope in 
2001.  We see no evidence for a bright companion 
in our NaCo images (see Figure \ref{f:mult}).
However, since  our AO images were taken 8 years later than the speckle data,
it is possible that the 
projected separation had changed enough in the intervening 
years for the binary to become unresolved.

Hence, our sample consists of eleven multiple systems, i.e. nine binaries (objects \# 1, 3, 4, 6, 7, 9, 14, 16, 17),
and two triples (objects \# 11 and 13).
Only in the cases of the close B/C pair in object \# 11  and \# 14 the 
binary
separation is small enough to suspect that the companions might  
have tidally disrupted the circumbinary disk thereby causing the inner hole 
inferred from the SED. However, in neither case the circumbinary nature can
be confirmed because it is unknown whether the IR excess in object \# 11 
originates in the wide A component or the close B/C pair
and the multiplicity of object \# 14 is not confirmed by our observations.  
We therefore only consider these two objects to be
circumbinary disk {\em{candidates}}. 
Table \ref{t:obs} lists the projected angular separations of the systems.

\subsubsection{Disk Masses}

\citet{andrews+williams05-1, andrews+williams07-1} modeled the 
IR and submillimeter SEDs of circumstellar disks and found 
a linear relation between the submillimeter flux and the disk mass
that has been calibrated by \citet{ciezaetal08-1} who obtained 
\begin{eqnarray}\label{eq:mass}
M_{\rm{DISK}} \sim8.0 \times 10^{-2}~[\frac{F_{\nu} (0.86~\rm{mm})} {\rm{mJy}} \times (\frac{d_{1}}{140~\rm{pc}})^2]~M_{\rm{JUP}},
\end{eqnarray}
where $d_{1}$ is the distance to the target. As described in Paper I,
disk masses obtained with the above relation are within a factor 
of  2 of model derived values, which is certainly good enough for the purposes of our survey.
However, one should keep in mind that larger systematic errors 
can not be ruled out \citep{andrews+williams07-1} as long as 
strong observational constraints on the grain size distributions and the
gas-to-dust ratios are lacking. 

Adopting distances of 150  pc to 
Lupus IV, 200  pc to Lupus III
\citep{comeron08-1}, 130 pc to Scp (Hatchell et al. in preparation), and 
150 pc to CrA \citep{Sicilia-Aguilaretal08-1} we use Equation~(\ref{eq:mass}) 
to estimate disk masses for the 17 systems 
in our sample (see Table \ref{t:derived}).
50$\%$ (i.e. 7/17) of the targets have been detected at 
8510 $\mu$m
(Table \ref{t:obs}). The corresponding disks masses range from $1-9$
\Mjup. Adopting a flux value of 3$\sigma$ for targets with non-detected
emission, we derive upper limits for the remaining targets of 
$\sim\,1-4$~\Mjup.  
Most of our targets have disk masses  $<~1-2$~\Mjup, 
but 5 targets have disk masses typical for CTTSs ($\sim$3 --
10~\Mjup). The most massive disks are detected around
\hbox{objects \# 2 and \# 3}, with 9 and 6~\Mjup, respectively.  

\subsubsection{Accretion rates}

The accretion rate is the second crucial 
parameter necessary to distinguish between 
the different mechanisms that may form inner opacity holes in 
circumstellar transitions disks. 
Most PMS stars show 
H$\alpha$ emission, either generated from 
chromospheric activity 
or magnetospheric 
accretion \citep{nattaetal04-1}. 
While non-accreting objects show rather narrow
($<200$ km s$^{-1}$) and symmetric line profiles of chromospheric
origin, the large-velocity magnetospheric accretion columns 
produce broad ($>200$ km s$^{-1}$) and asymmetric line profiles.  
As in Paper\,I we estimate the accretion rates of our transition disk 
systems according to the empirical relation obtained 
by \cite{nattaetal04-1}, i.e. 
\begin{equation}\label{eq:acc}
log(M_{\mathrm{acc}} (M_{\odot}~\rm{yr}^{-1})) =  -12.89  (\pm 0.3) + 9.7 (\pm 0.7) \times  10^{-3} \Delta V (\rm{km~s^{-1}})
\end{equation}
which is supposed to be relatively well calibrated for velocity 
widths covering \hbox{$\Delta\,V=200-600$~km s$^{-1}$}, 
which corresponds to accretion rates 
of $10^{-11}~\rm{M}_{\odot}~\mathrm{yr}^{-1}<\rm{M}_{\rm{acc}}<10^{-7}~
\rm{M}_{\odot}~\mathrm{yr}^{-1}$. However, the empirical dividing line between 
accreting and non-accreting objects has been placed slightly shifted by 
different authors at $\Delta\,V$ between 200 km s$^{-1}$ 
\citep{jayawardhanaetal03-1} 
and 270 km s$^{-1}$ \citep{white+basri03-1}. For systems with 
$\Delta\,V\sim\,200-300$~km s$^{-1}$ 
we therefore separate accreting and
non-accreting objects based on the (a)symmetry of the H$\alpha$ 
emission line profile and take into account the spectral type 
because accreting lower mass stars tend to have narrower 
H$\alpha$ emission lines.

To measure the H$\alpha$ velocity width 
$\Delta\,V$ we considered
for each system the spectral range that corresponds to
H$\alpha\pm 2500$ Km/s. The continuum plus emission profile were fitted
with a Gaussian plus parabolic profile. The parabolic fit was then used to
normalized the spectrum. 
A single Gaussian profile was sufficient here, being the emission 
single or double-peaked, as at this stage we were only interested 
in obtaining a good parabolic fit for the normalization.
Once the continuum had 
been normalized we measured $\Delta\,V$ at 10 per cent 
of the peak intensity.  
The obtained velocity dispersion 
of the H$\alpha$ emission lines 
and the corresponding accretion rate estimates are given in 
Table \ref{t:obs} and
Table \ref{t:derived}, respectively. 
The obtained accretion rates should be considered order-of-magnitude
estimates due to the large uncertainties associated with 
Equation~(\ref{eq:acc}) and the 
intrinsic variability of accretion in T\,Tauri stars.

\begin{figure}[t]
\begin{center}
\includegraphics[width=5in]{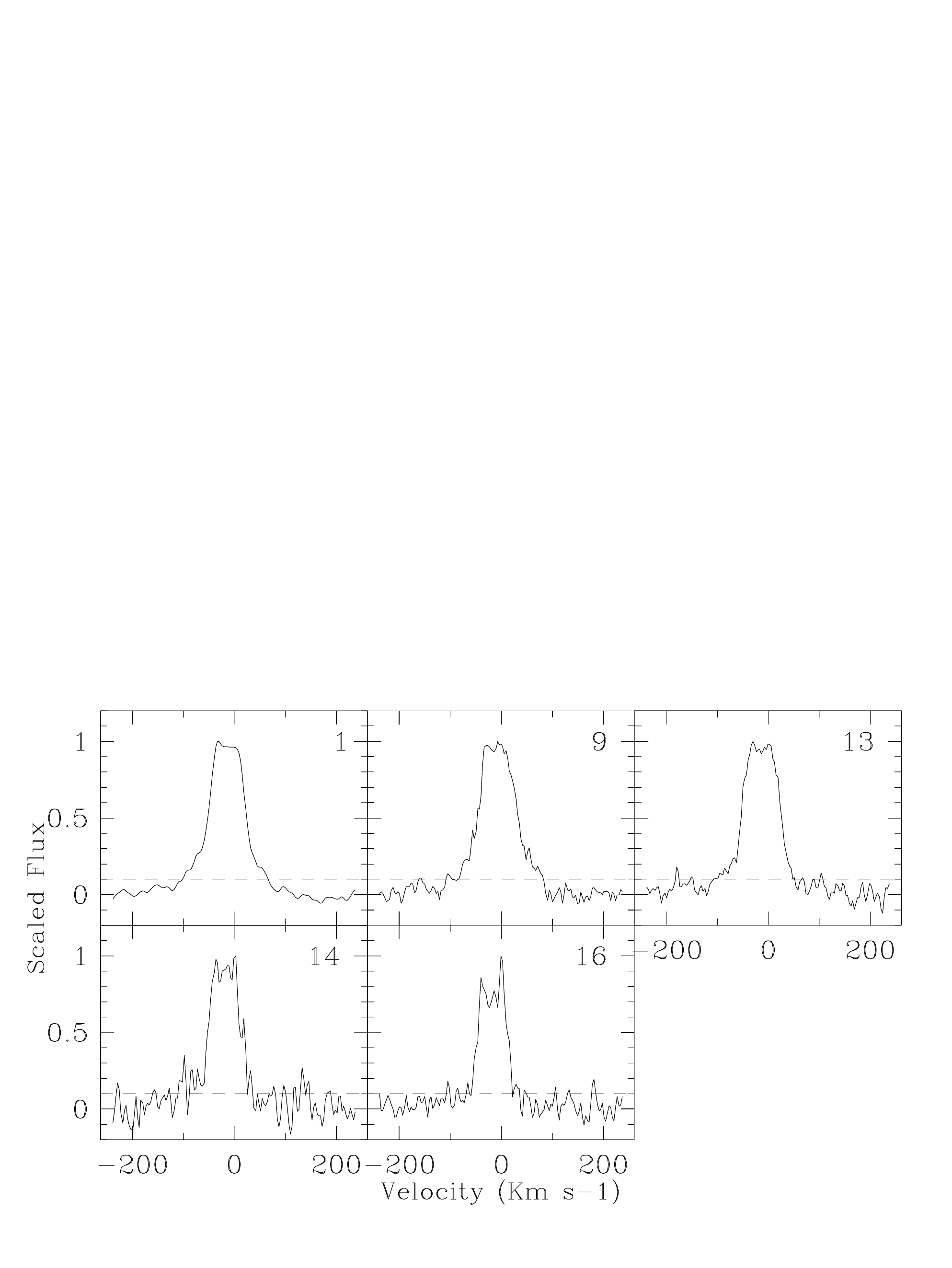}
\end{center}
\caption{\label{f:acc1}Normalized average H$\alpha$ profiles for non-accreting
  objects with clear H$\alpha$ emission. 
The horizontal dashed line indicates the 10\% peak intensity, where $\Delta$V is
measured. 
The velocity widths are $<$ 200 km s$^{-1}$ and the line
profiles are symmetric.}
\end{figure}

Our sample shows a large diversity of H$\alpha$ signatures. 
Five targets are classified as non-accreting objects that clearly show
symmetric and narrow H$\alpha$ emission line profiles 
($<$ 200 km s$^{-1}$, see Figure~\ref{f:acc1})
as expected from chromospheric activity.
For all these non-accreting objects,
we estimate an upper limit of the accretion rate, i.e.  
\hbox{$\rm{M}_{\mathrm{acc}}<10^{-11}$~M$_{\odot}$ yr$^{-1}$}. 

\begin{figure}[t]
\begin{center}
\includegraphics[width=5in]{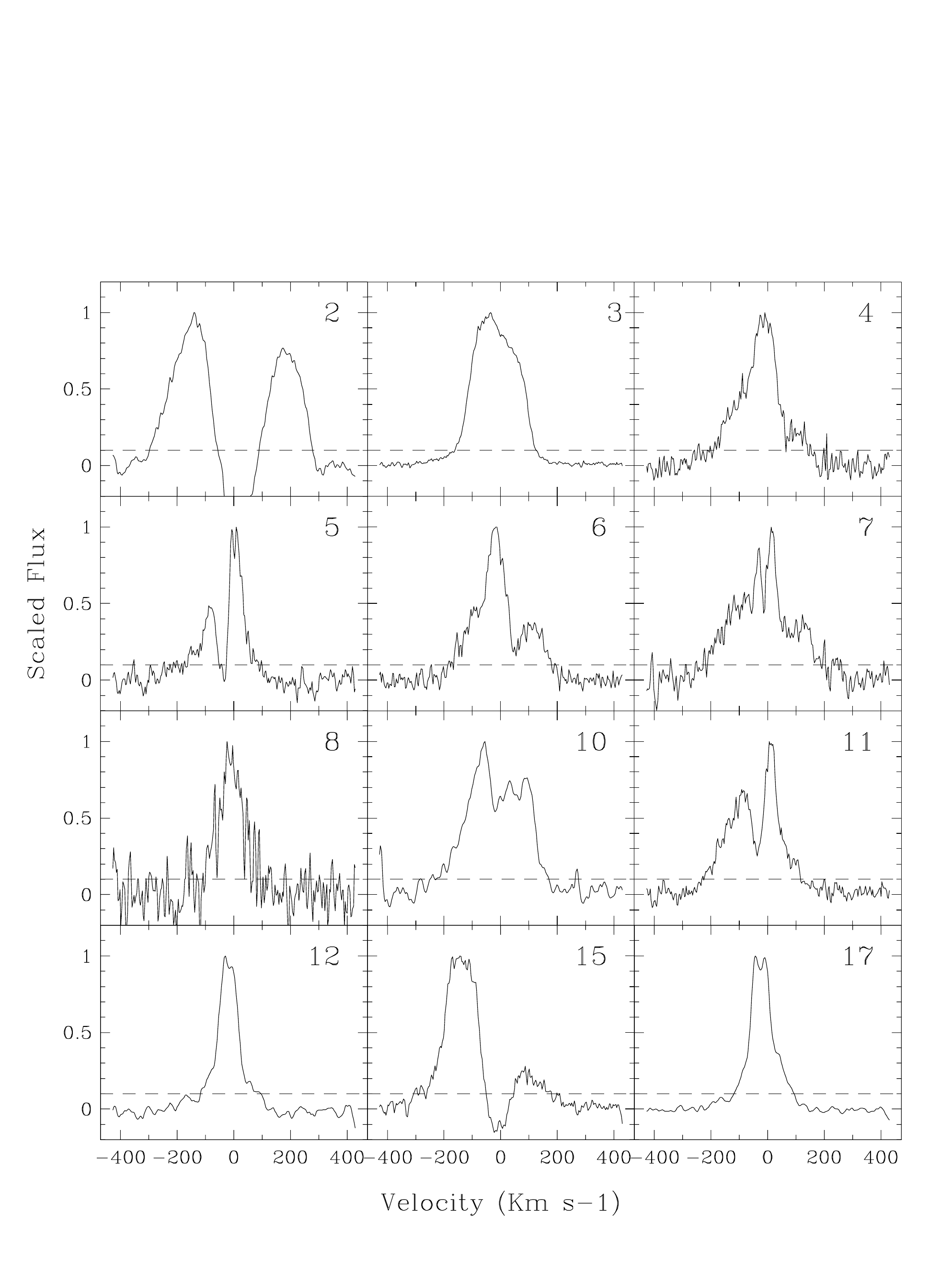}
\end{center}
\caption{\label{f:acc2}
Normalized H$\alpha$ profiles of the 12 accreting objects. 
These systems are considered accreting because either 
the velocity width is $>$ 200 km s$^{-1}$ or the line profile is asymmetric.
Note that objects \#\,8 and 12 represent borderline systems  
as the $\Delta$V$\sim200$km/s and the line is not clearly asymmetric
(see text for more details).}
\end{figure}

We classify the remaining 12 transition disk objects
as accreting. The majority of them (10) show 
clearly broad and asymmetric emission-line profiles 
(see Figure~\ref{f:acc2}). 
However, targets \# 8 and \# 12 
represent borderline cases with a rather small 
velocity dispersion for accreting systems 
($\Delta\,V\sim200$~\,km s$^{-1}$) and not clearly asymmetric 
line profiles. Such borderline systems require a more detailed discussion. 
Both stars are of late spectral type (M4-M5) and
very low-mass stars tend to have narrower H$\alpha$ lines than higher 
mass objects because of their lower accretion rates \citep{nattaetal04-1}
and their lower gravitational potentials \citep{muzzerolleetal03-1}. 
Object \# 12 additionally shows $8\,\mu$m excess emission 
indicating the presence of an inner disk. 
Given all the available data, we classify targets \# 8 and \# 12 as 
accreting objects, but warn the reader that their accreting nature 
is less certain than that of the rest of the objects classified as CTTSs. 
As accretion in T\,Tauri stars can well be episodic, mutli-epoch 
spectroscopy would be useful to unambiguously identify 
the accreting nature of these two systems. 

The mass accretion rates estimated for the 12 disks 
classified as accreting systems range
from \hbox{10$^{-11}$ M$_{\odot}$ yr$^{-1}$ to 10$^{-7.7}$ M$_{\odot}$
    yr$^{-1}$} 

\section{Discussion}

\subsection{The origin of the inner opacity hole}

With the collected information presented in the previous sections
we have at hand the following information of the {\em{Spitzer}}-selected 
transition disks in our sample: 

\begin{itemize}
\item Detailed SEDs that we quantify with the two-parameter scheme introduced by
\cite{ciezaetal07-1}, which is based on the longest wavelength at
which the observed flux is dominated by the stellar photosphere,
\lto \footnote{ 
To calculate \lto, we  compare the extinction-corrected  
SED with NextGen Models \citep{hauschildtetal99-1}
normalized to the J-band and choose \lto\,\,as the longest wavelength at 
which the stellar 
photosphere contributes over 50\% of the total flux. 
The uncertainty of \lto\,\,is roughly one SED point.}, 
and the slope of the IR excess, \afex, computed as
$d\log(F_{\lambda})/d\log(\lambda)$ between \lto ~and $24$\,$\mu$m.

\item Multiplicity information from the literature and AO IR
  imaging. 

\item Disk mass estimates based on measured submillimeter flux. 

\item Accretion rate estimates derived from H$\alpha$ line profiles. 

\end{itemize}

This information allows to 
separate the sample according to the physical processes that are the most
likely cause of the inner opacity hole: grain growth, planet formation,
photoevaporation, or close binary interactions. 
In what follows we briefly review each process that might be responsible for
the formation of the inner opacity holes, describe our criteria for
classifying transition disks, and discuss the corresponding 
sub-samples of transition disks obtained.  
 
\subsection{Accreting transition disks}

The presence of accretion in 
classical transition disk objects raises the question how 
the inner disk can be cleared 
of small grains while gas remains in the dust hole. 
The two mechanisms that can 
explain the coexistence of accretion and inner opacity holes are 
grain growth and dynamical interactions with (sub)stellar companions. 

\subsubsection{Grain growth}

Due to both, the higher densities and the 
faster relative velocities of particles in the inner
parts of the disk, this disk region offers much better conditions for 
dust agglomeration than the outer parts of the disk. Therefore, 
significant grain growth should start in the inner disk regions. 
As soon as the grains grow to sizes exceeding the considered wavelength 
($r>>\lambda$), the opacity 
decreases until an inner opacity hole forms. Early models by 
\citet{dullemond+dominik05-1} taking into account only dust coagulation 
predict much too short timescales of the order of $10^5$\,yrs to clear the
entire disk of small grains, which is inconsistent with observed SEDs of most
classical T\,Tauri stars. A more reliable picture combines 
coagulation and collisional fragmentation or erosion of large dust aggregates
\citep{dominik+dullemond08-1}. 

As a gradual transition between the inner and outer disk is predicted by grain
growth and dust settling models 
\citep{dullemond+dominik08-1,weidenschilling08-1}, 
grain growth dominated disks should have
$\alpha_{\mathrm{excess}}\le\,0$ (i.e., falling mid-IR SEDs)
while 
$\lambda_{\mathrm{turnoff}}$, associated with the size of the
hole, can differ over a rather wide range of values.  
Although grain growth does not directly affect the gas, 
it may increase accretion because the inner opacity hole 
can lead to efficient ionization and trigger the {\em{MRI}} instability
\citep{chiang+murray-clay07-1}. 

\begin{figure}[t]
\begin{center}
\includegraphics[width=5in]{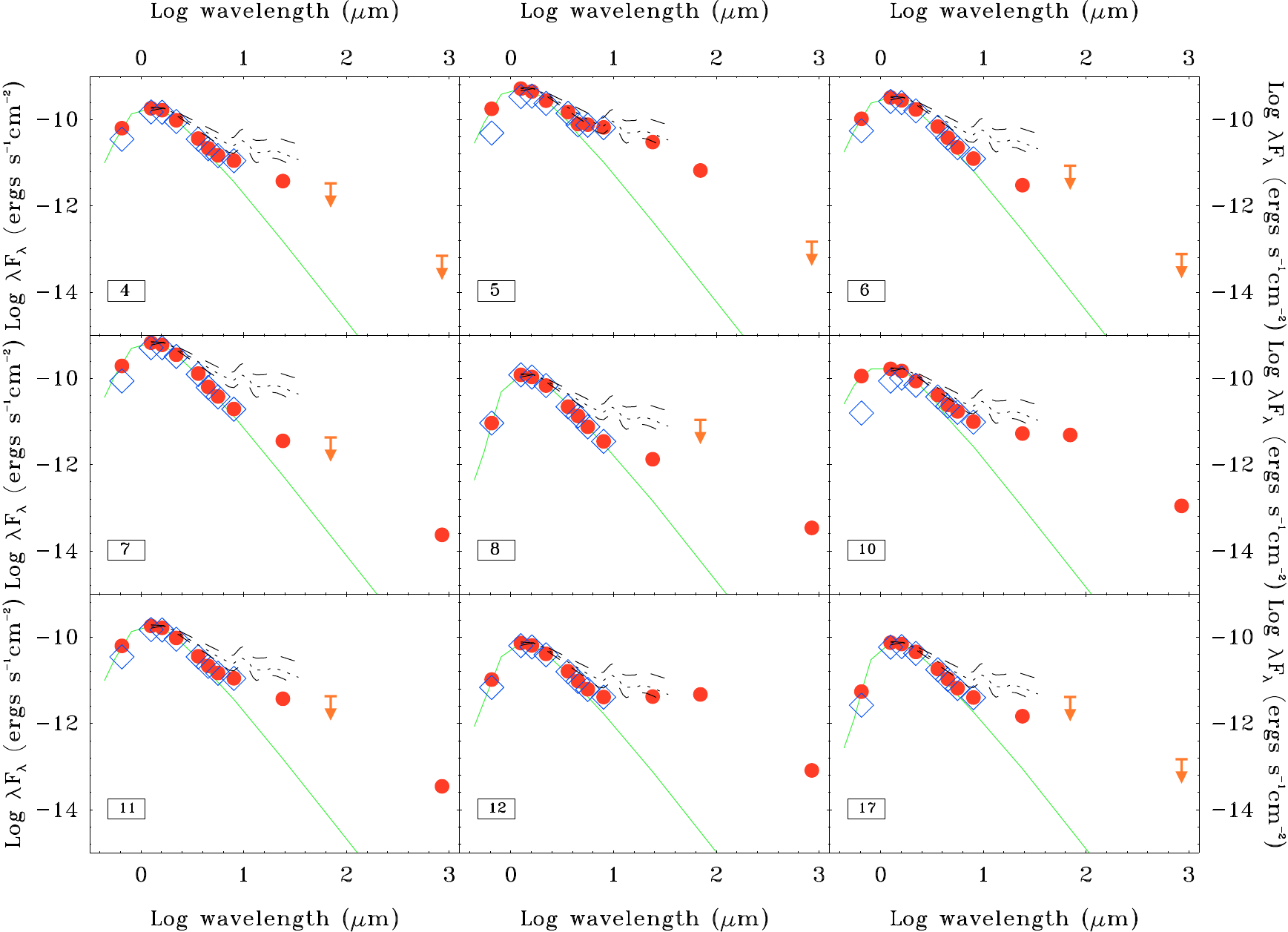}
\end{center}
\caption{SEDs of the 9 grain growth candidates. 
Disk masses range from $<$ 1.1 to 5.1~\Mjup, and accretion rates 
from $10^{-11}$ to $10^{-9.3}$ \hbox{\Msun~yr$^{-1}$}. The filled circles 
are detections, while the arrows represent 3$\sigma$ limits. 
The open squares correspond to the observed optical and near-IR fluxes 
before being corrected for extinction as in Paper~I.  
For each object, the average of the two R-band magnitudes (from the USNO-B1
catalog) listed in Table~\ref{t:mags} has been used. 
The classification of object \# 5 and 10 as grain growth dominated 
is slightly uncertain as classical T\,Tauris stars of late 
M-types can have similar SEDs \citep{ercolanoetal09-1}.
The solid line represents
the stellar photosphere normalized to the extinction-corrected J band. The
dotted lines correspond to the median mid-IR SED of CTTSs calculated by
\cite{furlanetal06-1}.
The dashed lines are the quartiles.
}
\label{f:grain-seds}
\end{figure}

Among the 17 transition disks in our sample, nine objects are accreting 
and are associated with \afex~$\leq\,0$. The corresponding SEDs are shown in
Figure \ref{f:grain-seds}. 
The grain growth candidate systems in our sample could be confounded with accreting classical T\,Tauri M-stars 
as predicted by \citet{ercolanoetal09-1}. 
However, most of our grain growth dominated disks 
have SEDs close to the stellar photosphere up to 
$\sim8-10\,\mu$m and we therefore do not expect significant 
contamination by non transition disks.
The only exceptions being objects \# 5 and 10 with a small value of 
\lto$=2.2$ and to some extend \# 4 and \# 11 (\lto$=4.5$). 
We recommend the reader to keep in mind
the uncertain classification of these four systems.

Compared with the Oph sample (Paper I) the accretion rates obtained for 
grain growth dominated disks  
are slightly smaller (i. e. $10^{-9.3}-10^{-11}$\Msun\,yr$^{-1}$).
This might be related to the slightly lower stellar masses or to 
advanced viscous evolution as discussed in \citet{dalessioetal06-1}.

The grain growth dominated disks are located
in the Lupus\, III, IV (8) and the CrA (1) star-forming regions.

\subsubsection{Dynamical interactions with (sub)stellar companions}

The truncation of the disk as the result of dynamical interactions with 
companions was first proposed by \cite{lin+papaloizou79-1}.
More recently, it has been shown that most PMS stars are in multiple systems 
with 
a lognormal semi-major axis distribution centered at $\sim\,30$\,AU  \citep[e.g.,][]{ratzkaetal05-1}. 
A significant fraction of the binaries in the star-forming regions considered
here should therefore be tight binaries with separations of $1-20$\,AU.
Disks in such close binary system will 
be tidally truncated at $\sim\,2~\times$ the binary separation 
and a circumbinary disk with an inner hole is formed 
\citep{artymowicz+lubow94-1}. The corresponding SED is that of a transition
disk. However, the circumbinary nature does not exclude additional 
evolutionary processes to be at work and we therefore 
provide an additional classification based on the disk properties only
(see Table \ref{t:derived}). 

Identification of companions that may cause the formation of a circumbinary
disk is possible either due to high-resolution imaging or by measuring radial
velocity variations. As described in Sect.~\ref{s:NaCo}, we identified 
2 circumbinary disk candidates among our 17 transition disk systems. 
One of them, object \# 11 was discovered by inspecting the NaCo images
obtained with the VLT, while the close binary nature of object \# 14
has been discovered by \citet{kohleretal08-1} using speckle interferometry 
at the NTT. 

Object \# 11 shows signs of strong accretion and has a SED 
with $\alpha_{\mathrm{excess}}\sim-1$ in agreement with grain growth.
It is currently not clear under which conditions 
gap-crossing streams can exist and allow accretion onto the central star to
proceed, but signs of accretion in circumbinary systems
\citep{carretal01-1,espaillatetal07-1} indicate that accretion is likely to
continue. On the other hand, object \# 14 is a non-accreting system 
such as the known binary CoKu Tau$/$4 \citep{ireland+kraus08-1}. 
According to its \Ld/\Ls~ratio we
classify this system  as a  circumbinary/photoevaporation disk
candidate. 

As a final note of caution, we would like to stress 
that both objects discussed above (\# 11 and \# 14) are 
circumbinary disk {\em{candidates}}. 
As all but one of our targets are M-type stars, most 
companions potentially responsible for their transition disk SEDs
are expected to lie at closer separations than those probed
by the AO images. Therefore, our sample of circumbinary 
disk candidates is incomplete and heavily biased towards large separations. 
Methods more sensitive to closer companions such as 
aperture masking and/or radial velocity observations are required 
to draw firm conclusions on circumbinary disk fractions.

\subsubsection{Giant Planet formation}

The most exciting way to produce a transition disk SED is by giant planet
formation. According to early models as well as recent numerical 
simulations, the formation of giant planets involves the formation of 
gaps and holes in the circumstellar disk
\citep{lin+papaloizou79-1,artymowicz+lubow94-1}.  

\begin{figure}[t]
\begin{center}
\includegraphics[width=5in]{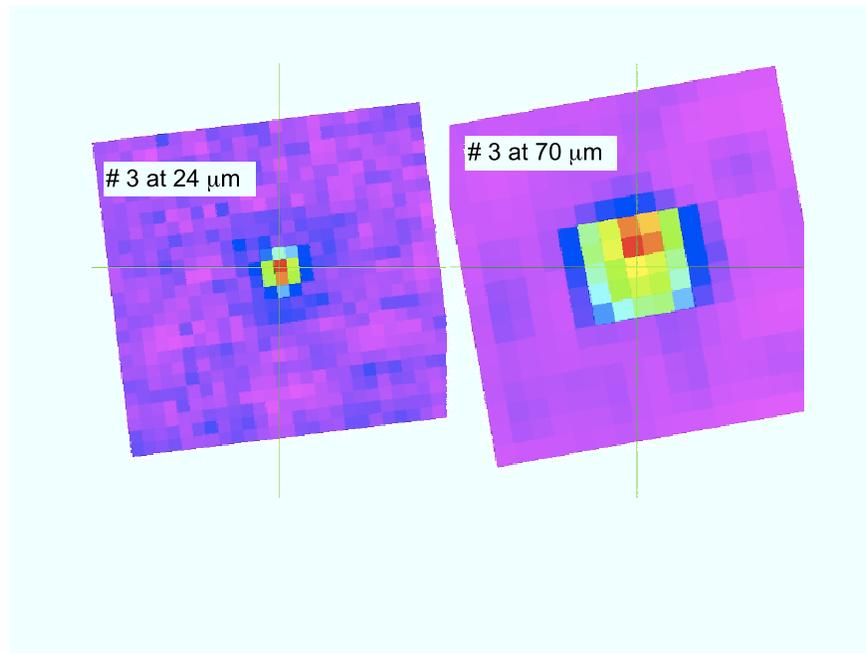}
\end{center}
\caption{
The 24 and 70 $\mu$m images  of object \# 3. We find no evidence 
for extended emission from the molecular cloud. In both mosaics, the detections 
are consistent with a point source at the location of the target
(marked by the crosshairs).
See the electronic edition of the Journal for a color version of this figure.
}
\label{f:tran6}
\end{figure}

As in the case of (sub)stellar companions it is uncertain if and to what extent 
accretion proceeds in the presence of 
a forming giant planet. Therefore, 
the most important sign of ongoing planet 
formation remains a sharp inner hole, usually corresponding to \afex$>0$
(i.e., a rising mid-IR SED). 
However, although very useful, the definition of \afex ~is incomplete, as the 
SED may also steeply rise at wavelengths longer than $24\,\mu$m. 
A spectacular example illustrating this is given 
by object \# 3. While ~\afex $\sim \,-2.2$, the
SED steeply rises between $24\,\mu$m and $70\,\mu$m. Furthermore, object
  \#\,3 shows clear signs of 
accretion ($M_{\mathrm{acc}}=~10^{-10.1}$~\Msun/yr) and of 
harboring a relatively massive disk (\Md$ ~\sim 6 $~\Mjup). 
Since this is a very atypical object, we verified that the large 70 $\mu$m
flux is not contaminated by extended emission from the molecular cloud. 
We examined the 24 and 70$\mu$m mosaics and verified that the detections 
are consistent with point sources at the source location (see
Figure~\ref{f:tran6}). 
A more typical transition disk system that might represent a currently 
planet forming disk is target \#\,15 with a clearly positive value of 
\afex ~and a high accretion rate.
A borderline case between grain growth and
planet forming disks is object \# 2. A high accretion rate combined with 
\afex$\sim0$ could be consistent with both scenarios. 
Keeping in mind the ambiguity, we classify 
object \# 2 as a planet forming disk candidate because it could potentially be 
an extremely interesting object. 
The SED of object \# 2 might be explained by a 
discontinuity in the grain size distribution rather than an inner opacity
hole. While the inner part of the disk 
still contains small grains, outer regions of the disk might be dominated by 
slightly larger dust particles. Such a scenario is in excellent 
agreement with the predictions of numerical simulations performed by
\cite{riceetal06-1}. They show that the 
planet--disk interaction at the outer edge of the gap cleared by a planet 
can act as a filter passed by small particles only which produces a
discontinuity in the dust particle size. 
To firmly establish its nature
object\,\# 2 deserves further follow-up observations
(e.g., high resolution imaging with ALMA).

\begin{figure}[t]
\begin{center}
\includegraphics[width=3.0in]{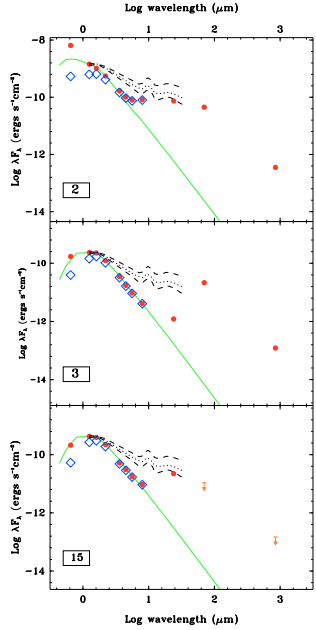}
\end{center}
\caption{SEDs of the three planet forming candidates. 
While object \# 15 can be considered a classical planet forming candidate 
system with \afex~$>0$ and \hbox{\lto~=~8.0~$\mu$m}, the other two systems are 
somewhat peculiar:
target \# 3 shows a very steep rise in flux observed at \hbox{70 $\mu$m}, which indicates a very large inner hole and \hbox{object \# 2} being relatively close to a full disk but with signs for a small and sharp inner hole. 
Disk masses are 9.1, 5.6 and \hbox{$<$ 2~\Mjup}~for objects \# 2, \# 3, 
and \# 15; respectively.  
The solid line as well as the dashes and dotted lines are the same as in
Fig.\,\ref{f:grain-seds}. 
}
\label{f:planet-seds}
\end{figure}

The SEDs of the three candidates for ongoing giant planet formation in our
sample are shown in Figure \ref{f:planet-seds}. 
The hosting forming giant planets candidates are located
in the Lupus III, IV (2) and the CrA (1) star-forming regions.

\subsection{Non-accreting objects}

The second main class of transition disks are those that do not show signs of
accretion. In such disks the inner opacity hole, i.e. the lack of small dust
particles in the inner disk regions, is likely to coincide with a gas hole,
i.e. the inner disk is completely drained.

\subsubsection{Photoevaporating disk}\label{sed_mor}

The most important process for clearing the inner disk 
in transitions disks that do not accrete 
is photoevaporation \citep[e.g.][]{alexanderetal06-1}. 
According to this model, extreme-ultraviolet (EUV) photons, originating in
the stellar chromosphere, ionize and heat the circumstellar hydrogen
which is then partly lost in a wind.
This process is supposed to work in all circumstellar disks but becomes
important only when the accretion rate drops to values similar to the
photoevaporation rate. Then, the inner disk drains 
on the viscous timescale supported by the generation of the {\em{MRI}}
\citep{chiang+murray-clay07-1}.  
Once an inner hole has formed, the inner disk rim is efficiently radiated 
and the entire disk should therefore quickly disappear. 
Photoevaporating disks should have negligible 
accretion \citep{williams+cieza11-1}. 
To separate photoevaporating disks from debris disks, we 
require the disk luminosity to be higher than that seen
in the brightest  bona-fide debris disks, i.e. \Ld/\Ls~$\geq\,10^{-3}$ 
\citep{brydenetal06-1,wyatt08-1}.
We thus obtained $70~\mu$m upper limits  from the noise of the  $70~\mu$m
{\em{Spitzer}} mosaics at the source location and calculated \Ld/\Ls~for  our
sample  
by integrating the stellar fluxes and disk fluxes over frequency (see Section
5.1.3 in Paper~I for  
details of the $70~\mu$m data analysis and the \Ld/\Ls ~calculation). 

\begin{figure}[t]
\begin{center}
\includegraphics[width=8.0cm]{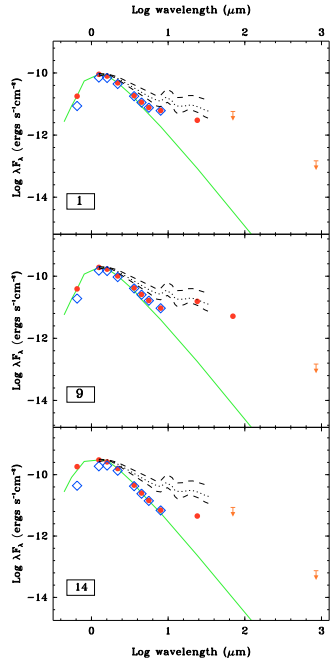}
\end{center}
\caption{SEDs of the 3 transition disks classified as photoevaporating 
disk 
candidates. The symbols, solid line as well as the dashes and dotted lines are
the same as in Figure~\ref{f:grain-seds}.
None of these systems have been detected at submillimeter wavelength and
consequently only upper limits for the disk masses could be derived. 
We conclude that
photoevaporation seems to be less efficient than has recently been suggested. 
}
\label{f:photo-seds}
\end{figure}

We classify three 
transition disks as photoevaporating disk candidates with 
negligible accretion (M$_{\mathrm{acc}}$ $<$ 10$^{-11}$
yr$^{-1}$) but \Ld/\Ls~$\geq\,10^{-3}$ (Table~\ref{t:derived}).
According to our submillimeter measurements, all these three systems 
have small disk masses ($<  1-3.4$~\Mjup, Table~\ref{t:derived}). 
In fact, for all photoevaporation candidates we could only derive upper
limits on the disk mass.
The SEDs of the three systems classified as photoevaporating disk objects 
are shown in Figure \ref{f:photo-seds}. 
The photoevaporated disks are located in the Lupus III, IV (2) and CrA (1)
star-forming regions.

\subsubsection{Debris disk}

\begin{figure}[t]
\begin{center}
\includegraphics[width=8cm]{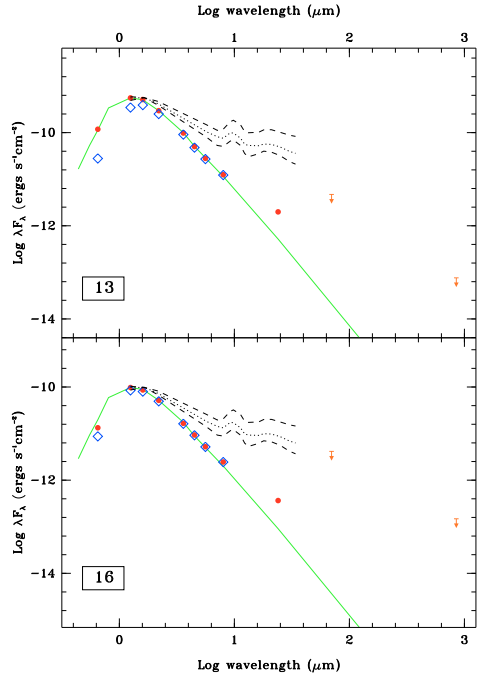}
\end{center}
\caption{SEDs of the 2 debris disk candidates identified in our sample. We distinguish between debris disks and 
photoevaporation applying an ad-hoc limit on \Ld/\Ls~$\lppr$~10$^{-3}$. The 
detection of the gas component in 
photoevaporating disks (e.g. with ALMA) may lead to a more physically 
motivated separation between the two sub-samples. 
The solid line as well as the dashes and dotted lines are the same as in Figure~\ref{f:grain-seds}.
\label{f:debris-seds}
}
\end{figure}

Photoevaporation can be considered as a transition stage 
between primordial and debris disks.
Debris disks contain a small amount of dust and are gas-poor. 
We find two debris disk candidates, i.e. non-accreting systems with
\Ld/\Ls$<\,10^{-3}$, among our 17 transition disks (see
Figure \ref{f:debris-seds}). 
The debris disks are located
in the CrA (1) and Scp (1) star-forming regions.

\subsection{Implications for disk evolution}

\subsubsection{Heterogeneity of transition disks}

\begin{figure}[t]
\begin{center}
\includegraphics[width=8cm]{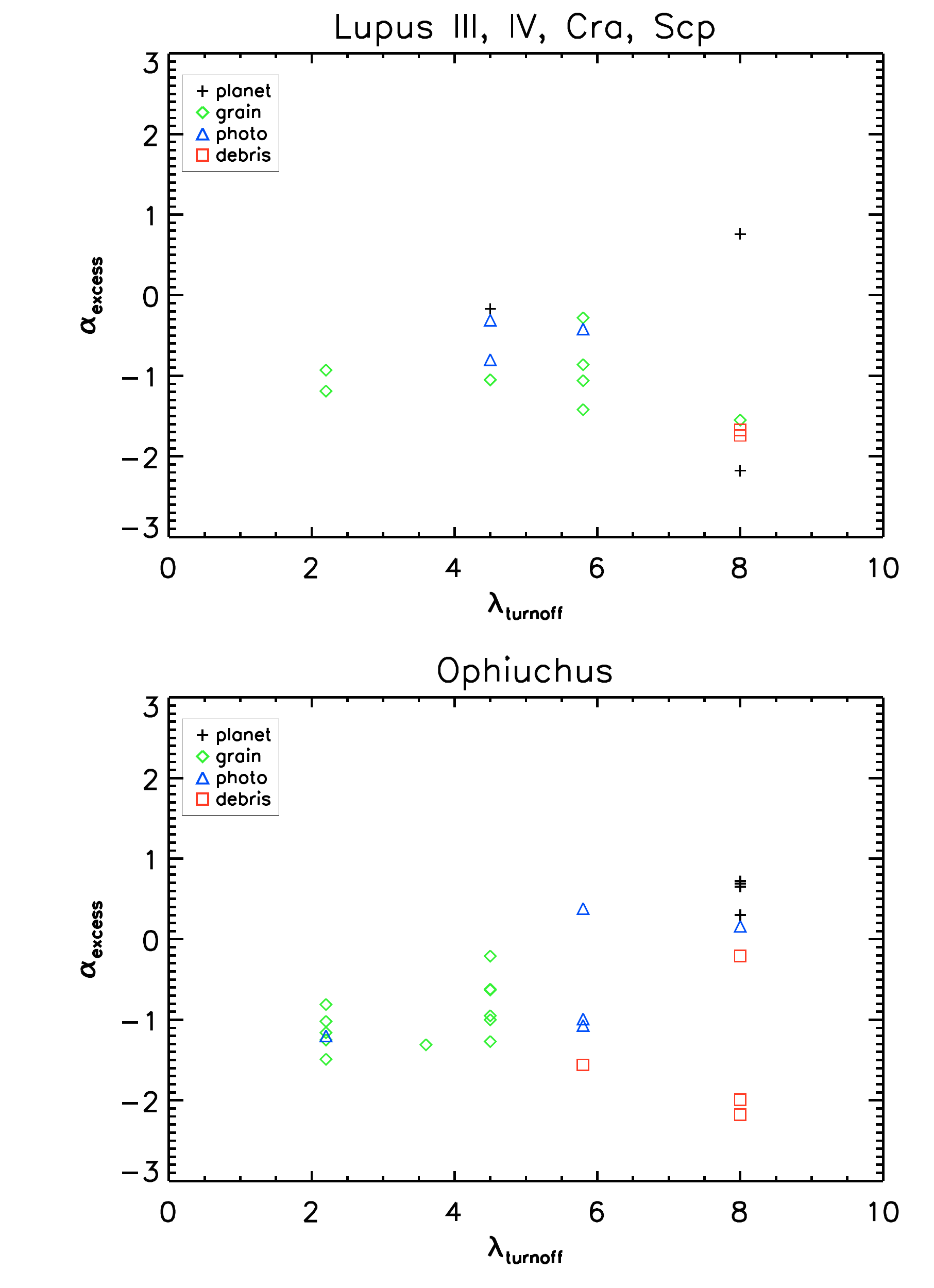}
\end{center}
\caption{
{\small
\afex~vs \lto~for transition disks identified in Paper\,I and the present work. 
The locations of the Ophiuchus transition disk sample from 
this work (top panel) and Paper\,I (bottom
panel) cover very similar ranges in the \afex~  versus \lto~plane. Different
symbols indicate different formation processes of the inner opacity hole. In
general, the Lupus, Cra, and Scp data confirm our previous findings: planet
forming disks have large values of \afex~  and \lto, grain growth dominated
disks should have small \afex~, but cover the entire range of \lto; and debris
disks have extremely low values of \afex~ and the IR excess starts at 
long \lto. However, two systems clearly show that \afex~and \lto~alone cannot fully characterize transition disks. We identified one planet forming disk candidate with \afex$<0$ but indications of a sharp hole at longer wavelength (object \# 3) and one planet forming candidate with \lto~=~4.5~$\mu$m (object \# 2) have been found.
}}
\label{f:alp-lam}
\end{figure}

In the previous sections, we presented detailed follow-up observations 
of 60 {\em{Spitzer}}-selected transition disk candidates located in the 
southern star-forming regions Lupus I, III, IV, V, VI,  CrA, and Scp. 
Optical spectroscopy revealed
that only 17 systems of these candidates 
are genuine transition disk T\,Tauri stars. Deriving
estimates for the accretion rates, disk masses, and multiplicity of 
these 17 systems we classified them as dominated by grain growth (9), 
giant planet formation (3), photoevaporation (3), or being in the final 
debris disk stage (2). Two of these transition objects, one grain growth (\# 11) and one photoevaporating (\# 14), are circumbinary
disk candidates, which offers the possibility of tidal truncation as mechanism responsible for an inner hole in the common/shared disk.
Combining these results with those presented in Paper\,I, 
we now have at hand well-defined and
well characterized samples of transition disks from several different 
star-forming regions. Figure~\ref{f:alp-lam} summarizes the properties of
these samples based on \afex~and \lto. 
The main aim of these series of papers is to 
progress with our understanding of circumstellar disk evolution 
and to compare transition disk samples of different clouds 
is key in this respect. 
Table~\ref{t:fractions} 
shows the fractions of different types 
of transition disks\footnote{Circumbinary disks are included twice in the 
table as binarity does not exclude a second process to cause the inner 
opacity hole in the disk. }
for Ophiuchus 
(age $\approx$ $0.3-2.1$ Myr, \citealt[][and references
  therein]{wilkingetal05-1}), CrA (age $\approx$ 1 Myr, 
\citealt{Sicilia-Aguilaretal08-1}), 
and Lupus I, III, IV (age $1.5-4$\,Myr, 
\citealt{merinetal08-1}). 
All YSO candidates followed up spectroscopically located in 
Lupus V, VI turned out to be AGB stars. 
These clouds have been recently 
estimated to be a few Myrs older 
\citep{spezzietal11-1} based on the dominance of 
Class III systems. As we have shown in 
Section~\ref{contamination}, at least $\sim50$\,\% of the claimed 
class III systems located in Lupus V, VI are very likely to be AGB stars.
This reduces the fraction of class\,III objects to values similar to 
those obtained for Lupus III. Based on this, Lupus V, VI, and III could well 
be of a very similar age.

The main result that can be obtained 
from Table~\ref{t:fractions} clearly is that 
young clouds ($\lappr\ 1-4$~Myr) contain a mixture of 
grain growth, photoevaporating, debris, and tidally disrupted transition 
disks. It is clear that all states of disk evolution are already present at 
this age range, which implies that different disks evolve at different rates 
and/or 
through different evolutionary paths. 
An important difficulty in constraining disk evolution 
is that stellar ages obtained from isochrones are very 
uncertain for individual systems. 
An analysis of the stellar age distributions of each 
disk category is therefore postponed to paper III 
(Cieza et al. 2012, ApJ submitted), where we discuss
a larger sample of well characterized transition disk 
objects including the systems presented here. 

\subsubsection{Evidence for low photoevaporation rates}

The general picture of photoevaporation is the following. 
In very young disks, the accretion rate largely exceeds
the evaporation rate and the disk evolves virtually unaffected by
photoevaporation. As the accretion rate is decreasing with time, 
the disk necessarily reaches the 
time when the accretion rate equals the photoevaporation rate and the outer
disk is no longer able to
resupply the inner disk with material. 
At this point, the inner disk drains on the 
viscous timescale ($\lappr$~$10^5$~\,yr) 
and an inner hole of a few AU in radius is formed
in the disk. The inner disk edge is now directly exposed to the EUV radiation
and the disk rapidly photoevaporates from the inside out. 

Early models of EUV photoevaporation predict evaporation rates of 
$10^{-10}-10^{-9}$\,\Msun/yr \citep{hollenbachetal94-1}. 
More recent simulations taking into account X-ray 
\citep{owenetal10-1} and/or
far-ultraviolet (FUV) irradiation \citep{gortietal09-1} in addition to 
the EUV photons,
largely exceed these early predictions, reaching
photoevaporation rates of the order of $10^{-8}$\,\Msun/yr 
\citep[see also][]{armitage11-1}.

As in steady state accretion disks  the  mass transfer through the disk  
is roughly proportional to the mass accretion rate onto the star,  
a crucial prediction of the 
photoevaporation model is that  high photoevaporation rates imply 
high disk masses at  the time the inner disk is drained. 
In particular,  models with  efficient  X-ray photoevaporation  predict 
a significant population of relatively massive ($\sim$7~\Mjup)
non-accreting transition disks \citep{owenetal11-1}. 

\begin{figure}[t]
\begin{center}
\includegraphics[width=5in,clip]{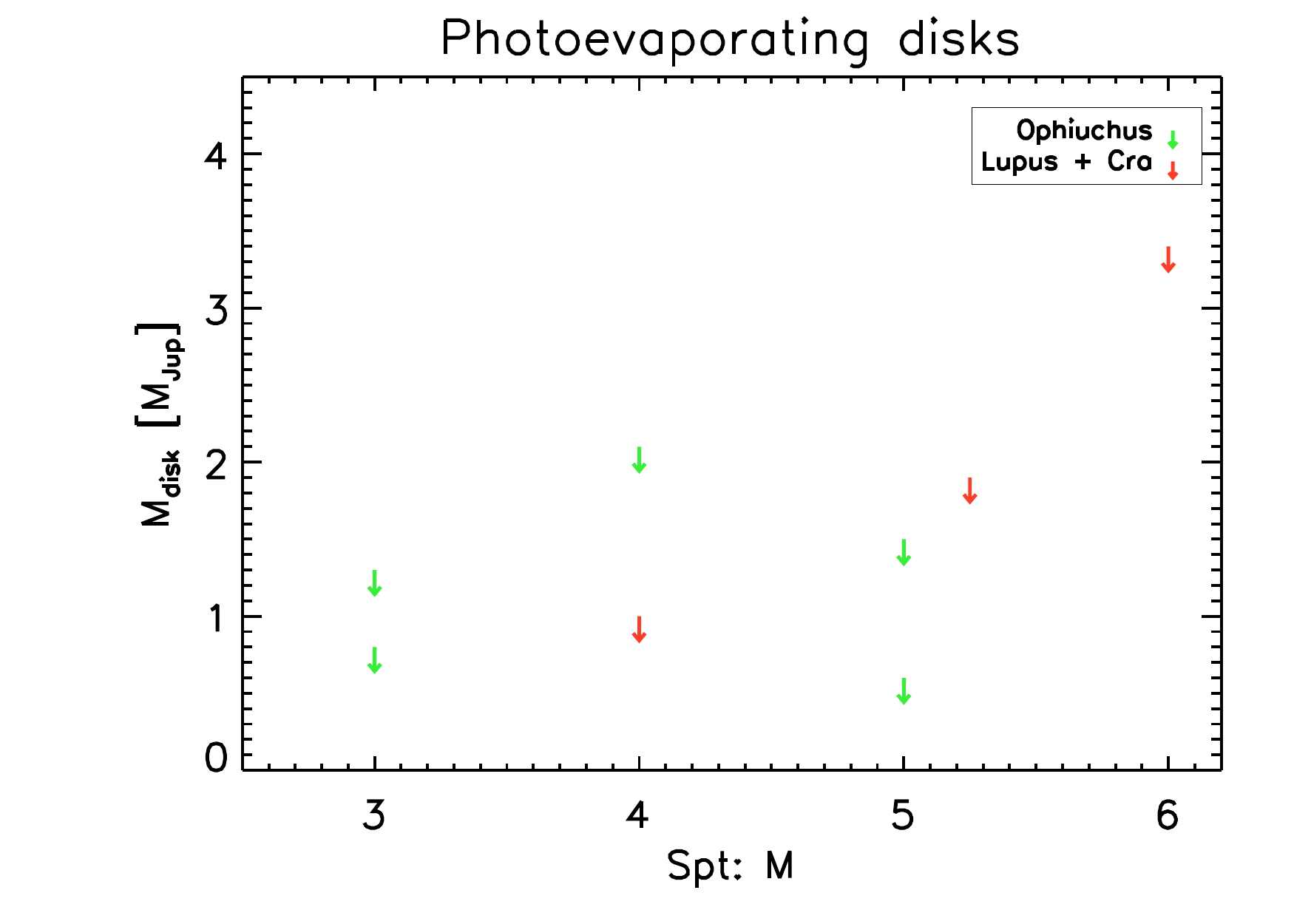}
\end{center}
\caption{The masses of the photoevaporating disk candidates in our sample. 
Clearly, massive photoevaporating disks predicted by high evaporation rates are absent.}
\label{f:photo-mass}
\end{figure}

Figure \ref{f:photo-mass} shows the upper limits (derived from submillimeter 
non detections) on the disk masses of the photoevaporating transition disks 
in all the clouds we
considered so far. 
Even taking into account uncertainties in our classification of 
photoevaporation 
candidates, it is evident that large numbers of non-accreting but massive 
disks do not exist. 
This indicates that photoevaporation is less efficient than predicted
by the models described above. However, 
one has to take into account that the sample of transition disks considered
here contains low-mass stars only while model calculations have been performed
exclusively for more massive stars $\sim 1$\Msun. Therefore, either 
a more homogeneous sample of photoevaporating disk systems 
covering a larger range of host star masses (earlier spectral types) 
or simulations of photoevaporation for disks around low-mass stars 
are required to provide a final answer on this issue.

\subsection{Current limitations and future perspectives}

Of course, our classification of transition disk objects is based on 
rather rough empirical relations and requires to carefully consider 
possible caveats.  
An obvious uncertainty concerns our multiplicity survey. 
The method of direct detection of companions is obviously more sensitive to 
binaries with large separations and low inclinations. 
Our NaCo observations are sensitive to 
projected separations of $\sim10-15$\,AU given the distance to our targets 
and -- depending on the intrinsic 
distribution of orbital separations -- we may therefore 
miss a significant fraction
of close binaries. To overcome this observational bias we are currently
performing radial velocity measurements of our targets using VLT/UVES. 
The method of detecting radial velocity variations is more sensitive to 
small separations and high inclinations and therefore  
complements the imaging results presented here.  
We will present the results in a forthcoming paper. 
However, the fact that only  6 of the 
43 transition disks studied herein
and in Paper~I are circumbinary disk candidates strongly suggests that 
binaries at the peak of their separation distribution ($\sim$ 30 AU) do not 
result in transition disk objects as such stellar  binaries would 
be easily detectable by our AO observations. Instead, they
are likely to destroy the disk rather quickly \citep{ciezaetal09-1}.

Another uncertainty in our classification procedure is the rather 
ad-hoc separation between photoevaporating and debris disk systems by 
using a limit in \Ld/\Ls. 
However, there is a physical and not only phenomenological 
difference between these two types of transition disks. 
Photoevaporating disks are dissipating 
primordial disks and should have gas rich outer disks 
while the debris disks should be 
gas poor. 
Molecular line observations with the
Atacama Large Millimeter/Submillimeter  Array 
(ALMA) of non-accreting disks will be able to distinguish between the two types of objects.

A huge problem related to the process of photoevaporation
is that the mass loss rates predicted by different models
differ by up-to two orders of magnitude 
\citep[see e.g.][]{williams+cieza11-1}. 
The disk mass at the time photoevaporation opens 
a hole in the disk is directly connected to the 
photoevaporation rates. Measuring the disk masses of
photoevaporating disks could therefore significantly constrain theoretical
models of photoevaporation. However, all of the photoevaporting disk candidates remain undetected and we can only put  upper limits to their masses. 
Fortunately, ALMA will be much more sensitive than all presently available
telescopes  
and will soon be able to measure the masses of many bona fide photoevaporating
disks. 
ALMA should also be able to measure, through  high resolution continuum observations at multiple wavelengths, 
the radial dependence in the grain size distribution expected in the 
grain-growth dominated disks. 

Finally, the recent identification using the aperture masking technique of 
what seems to be forming planets within the inner cavities of the transition 
disks around T Cha \citep{huelamoetal11-1} and LkCa 15 
\citep{kraus+ireland12-1} 
strongly encourages to obtain similar observations 
for the three planet-forming disk candidates identified herein,  
objects \# 2, \# 3, and  \# 15.
Any system with a planet still embedded in a primordial disk would represent 
an invaluable laboratory to study planet formation with current and future
instrumentation.

\section{Summary}

We have carried out a multifrequency study of {\em{Spitzer}}-selected YSO
transition disk candidates 
located in the Lupus complex (53), CrA (5), and Scp (2). 
We obtained submillimeter
observations (APEX), optical high resolution echelle
spectroscopy (Clay/Mike, Du Pont/echelle), and NIR 
images (from AO imaging VLT/NaCo). After deriving spectral types of each 
target, 43 AGB stars were removed (Lupus complex (41), CrA (1), and Scp (1)), 
leaving a sample of 17 genuine transition
disk  systems. We find that the vast majority of AGB 
stars have [3.6]-[24]  $<$ 1.8, underscoring the need for a spectroscopic
confirmation of YSO candidates with small 24$\mu$m excesses. 
The data obtained for the 17 transition systems
allows to estimate multiplicity, stellar accretion rates, 
and disk masses thereby allowing to identify the physical mechanism 
that is most likely to be responsible for the formation of the inner opacity 
hole. 
The observational results of this study can be summarized as follows:

\begin{enumerate}

\item The derived spectral classification indicates that all but one 
(object \# 2, K0) central star are
M-type stars, in agreement with previous results \citep{comeron08-1}.

\item 12/17 targets are accreting objects (i.e. asymmetric H$\alpha$ profile having a velocity width
\gappr ~200
km s$^{-1}$ at 10$\%$ of peak intensity). 

\item $\sim$ 50$\%$ of the sample are multiple systems and among them, two triple
systems. Two binary systems have small projected separations and are therefore candidates to host a 
circumbinary disk.

\item 7/17 targets have flux detection in the submillimeter. For the
remaining systems, we derive and upper limit of the disk mass (corresponding to a
flux of 3 $\times$ rms). The estimated disk masses for the detected objects 
cover the range 2~\Mjup--10~\Mjup.  

\end{enumerate}

Combining the derived estimates of disk masses, 
accretion rates and multiplicity with the 
SED morphology and fractional disk luminosity 
(\Ld/\Ls) allows  to classify the disks as strong
candidates for the following categories:
\begin{itemize}

\item 9/17  grain growth-dominated disks  (accreting
objects with negative SED slopes in the mid-IR, \afex $<$ 0).

\item 3/17  photoevaporating disks (non-accreting objects
with disk mass  $<$ 3~\Mjup, but \hbox{\Ld/\Ls $>$ 10$^{-3}$}). 

\item 2/17  debris disks (non-accreting objects with disk
mass  $<$ 2.1~\Mjup~and \hbox{\Ld/\Ls $<$ 10$^{-3}$}).

\item 2/17  circumbinary disks (a binary tight enough to
accommodate both components within the inner hole).
 
\item 3/17 giant planet forming disk (accreting systems with SEDs 
indicating sharp inner holes).

\end{itemize}

Inspecting in more detail the different sub-clouds 
analyzed in this study we find
the same heterogeneity of the transition disk population in Lupus 
III, IV, CrA as in our previous analysis of transition disks in Ophiuchus 
\citep[][Paper\,I]{ciezaetal10-1}. 
We therefore conclude that 
photoevaporation, giant planet formation, and grain growth
produce inner holes on similar timescales.
Not a single transition disks has been found in Lupus I, V, VI. All 33 candidates that have been spectroscopically followed up  
turned out to be AGB stars which questions the recent interpretation of 
\citet{spezzietal11-1} that  Lupus I, V, VI might be relatively 
old star forming regions dominated by Class\,III objects. 

In addition, our detailed observational analysis of transition disks
provides clear constraints on theoretical models of disk photoevaporation
by the central star. 
According to the large  evaporation rates predicted by 
recent models \citep[i.e. see][]{armitage11-1},  
large numbers of massive photoevaporating 
transition disks systems should exist. 
In contrast to this prediction, all photoevaporating 
disk candidates identified in this 
work and Paper\,I contain very little mass, indicating  
much smaller evaporation rates at least for the low-mass stars
considered here. 
Similarly, the low incidence of circumbinary transition 
disk candidates ($\sim$ 10$\%$)  supports the idea that  most 
disks are destroyed rather quickly by companions at $\sim 10-40$ AU
separations.  

Finally, we emphasize that the 43 transition disk systems discussed in
this work and in Paper\,I represent the currently largest and 
most homogeneous sample of 
well-characterized transition disks. 
Further investigating these systems with new observing capabilities such as 
ALMA therefore holds the potential to significantly improvement our 
understanding
of the physical processes driving circumstellar disk evolution.

\acknowledgments
{\it{Acknowledgments:}} 
GAR was supported by 
ALMA/Conicyt (grant 31070021) and ESO/comit\'e mixto.  
MRS acknowledges support from Millennium Science Initiative, 
Chilean ministry of Economy: Nucleus P10-022-F 
and Fondecyt (grant 1100782). 
LAC acknowledges support provided by NASA through the {\em{Sagan}}
Fellowship Program. ARM thanks for financial support from  Fondecyt in the form of grant
number 3110049,  ESO/comite mixto and Gemini/Conicyt (32080023).
ASC was supported by grants from Consejo Nacional de
Investigaciones Cient\'{\i}ficas y T\'ecnicas de la Rep\'ublica
Argentina, Agencia Nacional de Promoci\'on Cient\'{\i}fica y Tecnol\'ogica and
Universidad Nacional de La Plata (Argentina).
We finally thank Dr. Giorgio Siringo and Dr. C. De Breuck for assistance with 
performing the APEX observations and the corresponding data reduction.
We are also grateful for the support of the staff at Las Campanas Observatory. 
A special thank is given to  Dr. M. Orellana and Evelyn Puebla for their 
help during the first Las Campanas observing run.
APEX is a collaboration between the Max-Planck-Institut fur
Radioastronomie, the European Southern Observatory, and the Onsala Space
Observatory. This work makes 
use of data obtained with the {\em{Spitzer}} Space Telescope,
which is operated by JPL/Caltech, under a contract with NASA.
{\em{Facilities}}: {\em{Spitzer}} (IRAC, MIPS), VLT:Yepun, Magellan: Clay, Du
Pont (Echelle)

\bibliographystyle{apj}

%\bibliography{aamnem99,tdbib}

%\include{figslinda}

\begin{landscape}
\begin{deluxetable}{rrcrrrrrrrrrrrcc}
%\rotate
\tablewidth{0pt}
%\tablewidth{0pt}
%\tablewidth{\textwidth}
\tabletypesize{\tiny}
\tablecaption{{\bf Spectroscopically identified AGB stars}}
\tablehead{\colhead{\#}&\colhead{Ra (J2000)}&\colhead{Dec (J2000)}&\colhead{\emph{Spitzer} ID}&\colhead{R}&\colhead{J\tablenotemark{a}}&\colhead{H}&\colhead{K$_{S}$}&\colhead{F$_{3.6}$\tablenotemark{a}}
&\colhead{F$_{4.5}$}&\colhead{F$_{5.8}$}&\colhead{F$_{8.0}$}&\colhead{F$_{24}$}&\colhead{Region} &\colhead{References\tablenotemark{b}} \\
\colhead{}&\colhead{(deg)}&\colhead{(deg)}&\colhead{}&\colhead{(mag)}&\colhead{(mJy)}&\colhead{(mJy)}&\colhead{(mJy)}&\colhead{(mJy)}&\colhead{(mJy)}&\colhead{(mJy)}&\colhead{(mJy)}&\colhead{(mJy)}&&}
\startdata
1 & 234.51292 & -33.23269   &    SSTc2d\_J153803.1-331358 & 13.35         &3.91e+02 &   6.91e+02&  6.83e+02 &  3.47e+02& 2.20e+02 & 1.82e+02 & 1.26e+02     & 3.56e+01  & Lup I  & 1    \\
2 &  235.64750 & -34.37292    & SSTc2d\_J154235.4-342223 & 14.17          & 5.16e+02 &  8.86e+02&  8.43e+02 &  4.00e+02& 2.66e+02 & 2.03e+02 & 1.40e+02     & 4.96e+01  & Lup  I  &   \\
 3 &  239.93868   &   -41.91590  &    SSTc2d\_J155945.3-415457 & 13.18  &  3.64e+02  &   7.92e+02 & 8.86e+02 &  5.78e+02 &  3.35e+02  & 3.23e+02 &  3.80e+02 & 2.68e+02 &Lup IV &1 \\
 4 &  240.37369   &   -42.13432  &    SSTc2d\_J160129.7-420804 & 15.74  &  7.74e+01  &   1.57e+02 & 1.60e+02 &  9.11e+01 &  5.86e+01  & 4.89e+01 &  3.80e+01 & 1.79e+01 &Lup IV  & 1\\
 5 & 240.62463 &  -41.85307       & SSTc2d\_J160229.9-415111   &  17.91 &  3.95e+01  &   8.07e+01  &8.73e+01 &  4.99e+01 & 3.01e+01 & 2.55e+01 & 1.82e+01 &    5.15e+00  &    Lup  IV &1\\
 6 &  242.19953   &   -38.83361  &    SSTc2d\_J160847.9-385001 &13.71  &   1.12e+03  &   1.93e+03 & 1.89e+03 &  9.86e+02  & 4.50e+02   &4.37e+02  & 3.14e+02 & 1.41e+02& Lup III   &  \\
 7 &242.39212 & -39.22835          &SSTc2d\_J160934.1-391342 & 15.11    &   5.36e+02 &    1.20e+03 & 1.43e+03 &  5.33e+02&   3.99e+02&  4.66e+02&  2.83e+02 & 5.35e+01&    Lup  III & 1\\
 8 &  242.50045   &   -38.90031  &    SSTc2d\_J161000.1-385401 &17.47  &   2.25e+02  &    4.13e+02 &5.55e+02 &  3.66e+02  & 1.76e+02   &2.68e+02  & 2.06e+02 & 9.72e+01 &Lup III   & 1 \\
 9 &  242.85827   &   -39.18979  &    SSTc2d\_J161126.0-391123 & 16.29  &  6.96e+02  &   1.43e+03&  1.71e+03&   1.01e+03 &  5.99e+02  & 5.56e+02 &  3.62e+02 &  1.05e+02& Lup III &1\\
 10 &  243.21550   &   -38.70443  &    SSTc2d\_J161251.7-384216 &14.06  &
 5.97e+02  &   1.04e+03 & 1.07e+03 &  6.37e+02  & 3.49e+02   &3.06e+02  &
 2.02e+02 & 7.43e+01 &Lup III  & 1 \\
11 & 244.92350 &  -37.78246        &SSTGBS\_J16194163-3746568  & 16.36   &  6.48e+01  &   1.27e+02 & 1.33e+02 &  8.56e+01&   5.64e+01&  4.77e+01&  3.26e+01        &  1.02e+01&   Lup  V &2    \\
 12 &  245.00856   &   -41.62400  & SSTGBS\_J16200205-4137264 &15.37  &  6.17e+02  &   1.24e+03 & 1.35e+03 &  7.27e+02  & 4.28e+02   &3.60e+02  & 2.31e+02 & 9.00e+01 &Lup VI  & 2 \\
 13 &  245.03960   &   -41.43343  &  SSTGBS\_J16200950-4126003 &15.78  &  7.69e+02  &   1.77e+03 & 2.05e+03 &  1.40e+03  & 8.77e+02   &7.88e+02  & 5.20e+02 & 1.61e+02 &Lup VI   & 2 \\
 14 &  245.13167   &   -37.51133  & SSTGBS\_J16203160-3730407 &16.96  &  1.40e+03  &   2.78e+03 & 3.38e+03 &  2.07e+03  & 1.24e+03   &1.27e+03  & 9.10e+02 & 2.11e+02 &Lup V    & 2 \\
 15 &   245.22704 & -36.91195         & SSTGBS\_J16205449-3654430 &  16.03 & 1.89e+02 &    3.86e+02 & 4.12e+02 &  2.24e+02 &  1.16e+02&  9.97e+01& 6.43e+01       &  2.51e+01&   Lup  V& 2 \\
 16  &  245.35836  &  -37.51710   &   SSTGBS\_J16212600-3731015 &16.69   & 1.62e+02   &  4.21e+02 & 6.41e+02  & 7.57e+02 &  5.29e+02  & 5.18e+02 &  3.87e+02 &  1.61e+02& Lup V   & 2  \\
 17 &   245.37467       & -37.03926  &      SSTGBS\_J16212991-3702213 & 14.84& 1.68e+02  &   3.15e+02 & 3.31e+02 &  1.89e+02 & 9.94e+01 & 8.37e+01 & 5.79e+01    &  5.69e+01 &Lup  V &2 \\
 18  &   245.41990   &  -41.37274   &    SSTGBS\_J16214077-4122218 &15.98 &  1.26e+02  &   2.51e+02&  2.64e+02 &  1.58e+02 &  8.63e+01 &  7.36e+01 &  4.86e+01 &  1.92e+01 &Lup VI & 2\\
19  &    245.45722  &   -41.11716  &     SSTGBS\_J16214973-4107017 &15.35 & 1.20e+0  &    2.30e+02&  2.65e+02 &  1.92e+02 &  1.34e+02 &  1.08e+02 &  7.48e+01 &  2.40e+01 &Lup VI &2\\
 20 &    245.48435  &   -37.22131  &     SSTGBS\_J16215624-3713167 &17.18 & 1.95e+0  &    3.89e+02 & 4.34e+02 &  2.43e+02 &  1.31e+02 &  1.15e+02 &  7.38e+01 &  2.84e+01 &Lup V &2 \\
 21 &    245.50680  &   -37.27693  &     SSTGBS\_J16220163-3716369&16.20 & 5.48e+0  &    1.06e+02 & 1.09e+02 &  6.50e+01 &  4.23e+01 &  3.53e+01 &  2.48e+01 &  8.24e+00 &Lup V &2\\
 22 &    245.57457  &   -36.97127  &    SSTGBS\_J16221789-3658165&16.80  &  2.68e+02  &   5.24e+02 &5.83e+02 &  3.42e+02  & 1.95e+02  & 1.68e+02  & 1.08e+02  & 3.44e+01 &Lup V &2\\
 23 &    245.63924  &   -41.05499  &    SSTGBS\_J16223341-4103179 &16.34  &  5.72e+02  &   1.32e+03 &1.50e+03 &  8.22e+02  & 4.95e+02  & 4.26e+02  & 2.87e+02  & 1.21e+02 &Lup VI&2 \\
 24 &    245.67774  &   -37.35645  &    SSTGBS\_J16224265-3721232&16.08  &  1.97e+02  &   4.34e+02 &4.97e+02 &  3.61e+02  & 2.74e+02  & 2.47e+02  & 2.61e+02  & 1.98e+02 &Lup V &2\\
 25 &    245.76534  &   -37.49477  &    SSTGBS\_J16230368-3729411 &16.02  &  6.05e+01  &   1.18e+02 &1.20e+02 &  6.64e+01  & 4.17e+01  & 3.55e+01  & 3.12e+01  & 1.86e+01 &Lup V &2 \\
 26 &       245.79592 & -41.28620 &      SSTGBS\_J16231101-4117103&16.48  & 1.14e+02   &   2.64e+02 & 3.00e+02 & 1.67e+02   & 1.15e+02 & 9.21e+01   & 6.16e+01  & 1.79e+01 &Lup  VI & 2\\
 27 &  245.81152  &   -40.28753  &    SSTGBS\_J16231476-4017151 &16.96  &  1.69e+02  &     3.56e+02 &4.07e+02 &  2.20e+02  & 1.36e+02  & 1.15e+02  & 8.15e+01  & 2.86e+01 &Lup VI &2\\
 28 &  245.81661  &   -41.06693  &    SSTGBS\_J16231598-4104009 &13.11  &  1.97e+03  &     4.16e+03 & 4.86e+03 & 1.80e+03  & 1.31e+03  & 1.12e+03  & 7.36e+02  & 2.41e+02 &Lup VI &2\\
 29  & 245.85994   &  -37.86896   &   SSTGBS\_J16232638-3752082 &17.31  &  1.20e+02   &    2.35e+02 & 2.99e+02 & 3.24e+02  & 2.47e+02  & 2.33e+02  & 1.74e+02  & 4.96e+01 &Lup V &2\\
 30  & 245.88526   &  -37.87672   &   SSTGBS\_J16233246-3752361 &15.74  &  4.72e+02   &    9.78e+02 & 1.06e+03 & 5.68e+02  & 3.23e+02  & 2.85e+02  & 1.92e+02  & 7.09e+01& Lup V &2\\
 31 &  245.91108  &   -37.52745  &    SSTGBS\_J16233865-3731388 &15.80  &  1.27e+02  &     2.39e+02 & 2.53e+02 & 1.43e+02  & 7.59e+01  & 7.05e+01  & 5.16e+01  & 3.14e+01 &Lup V & 2\\
 32 &  245.95431  &   -40.43823  &    SSTGBS\_J16234903-4026176 &17.12  &  9.57e+01  &     2.08e+02 & 2.33e+02 & 1.64e+02  & 1.13e+02  & 9.72e+01  & 7.79e+01  & 2.52e+01 &Lup VI &2\\
 33 &  245.99615  &   -37.89800  &    SSTGBS\_J16235907-3753528 &14.28  &  2.23e+03  &     4.60e+03 & 5.14e+03 & 2.51e+03  & 1.61e+03  & 1.41e+03  & 9.50e+02  & 3.19e+02 &Lup V & 2\\
 34 &  246.11110  &   -37.96131  &    SSTGBS\_J16242666-3757407 &17.52  &  3.13e+02  &     6.67e+02 & 7.63e+02 & 4.10e+02  & 2.14e+02  & 2.02e+02  & 1.29e+02  & 5.62e+01 &Lup V & 2\\
 35 &  246.23294  &   -40.19118  &    SSTGBS\_J16245590-4011282 &15.15  &  1.06e+03  &     2.09e+03 & 2.65e+03 & 1.43e+03  & 9.49e+02  & 9.54e+02  & 6.74e+02  & 1.78e+02 & Lup VI& 2\\
 36 &  246.27875  &   -38.05595  &    SSTGBS\_J16250690-3803214 &14.99  &  1.34e+03  &     2.71e+03 & 2.98e+03 & 1.64e+03  & 9.41e+02  & 8.20e+02  & 5.39e+02  & 1.78e+02 &Lup V & 2\\
 37 &  246.46860  &   -40.31346  &    SSTGBS\_J16255246-4018484 &17.39  &  1.23e+02  &     2.63e+02 & 2.93e+02 & 1.23e+02  & 7.96e+01  & 7.38e+01  & 4.68e+01  & 1.28e+01 &Lup VI &2\\
 38 &  246.49321  &   -40.16614  &    SSTc2d\_J16255837-4009581 &16.88  &  5.17e+02  &     1.20e+03 & 1.44e+03 & 7.77e+02  & 4.23e+02  & 4.06e+02  & 2.63e+02  & 1.05e+02 &Lup VI& 2\\
 39 &  246.55582  &   -39.83173  &    SSTc2d\_J16261339-3949542 &14.83  &  1.68e+02  &     3.01e+02 & 3.55e+02 & 2.15e+02  & 1.60e+02  & 1.34e+02  & 1.01e+02  & 3.70e+01 &Lup VI& 2\\
 40  & 246.60631   &  -39.74646   &   SSTGBS\_J16262551-3944472 &15.44  &  4.76e+01   &    9.31e+01 & 9.09e+01 & 4.70e+01  & 3.20e+01  & 2.52e+01  & 2.94e+01  & 1.53e+01 &Lup VI &2\\
 41  & 246.96060   &  -39.80278   &   SSTGBS\_J16275054-3948100 &16.20  & 3.15e+02   &     6.61e+02 & 7.35e+02 & 4.71e+02  & 2.81e+02  & 2.39e+02  & 1.59e+02  & 5.74e+01 &Lup VI &2\\
 42 &  252.27333       & -15.62027   & SSTc2d\_J16490560-1537129 &17.92
 &1.40e+02 & 6.63e+02 & 1.12e+03 & 1.00e+03 & 7.78e+02 & 7.42e+02 & 5.17e+02 & 1.46e+02 & Scp & 4 \\
 43 &   285.78812   &   -36.95611  & SSTGBS\_J19030915-3657220    &17.45  &  6.88e+01   &   1.97e+02 & 2.43e+02 & 1.63e+02  &1.21e+02   &1.01e+02   & 6.75e+01   & 2.04e+01 &Cra & 3
\enddata
\tablenotetext{a}{All the 2MASS, IRAC, and 24 $\mu$m detections are $\ge$ 7$\sigma$ (i.e., the photometric uncertainties
are $\lesssim$ 15$\%$)}
\tablenotetext{b}{References of previous works that cataloged the target as YSOc: (1) \citet{merinetal08-1}; (2) \citet{spezzietal11-1}; (3) \citet{petersonetal11-1}}
\label{t:giants}
\end{deluxetable}
\end{landscape}

\begin{deluxetable}{cccccccc}
\tabletypesize{\scriptsize}
\tablewidth{0pt}
\tablecaption{\label{t:cont}{\bf T\,Tauri and AGB star fractions}}
\tablehead{\colhead{Cloud}&\colhead{Coordinates}&\colhead{Sample}&\colhead{Transition Disk}&\colhead{AGB}&\colhead{Percentage}&\colhead{Age}&\colhead{Cloud's}\\
\colhead{}&\colhead{(l,b)}&\colhead{}&\colhead{candidates}&\colhead{}&\colhead{of AGB}&\colhead{[Myr]}&\colhead{Distance}\\
\colhead{}&\colhead{$\approx$ (deg,deg)}&\colhead{\#}&\colhead{\#}&\colhead{\#}&\colhead{$\%$}&\colhead{}&\colhead{[pc]}}
\startdata      
Lupus I        &      339, 16  &     2       &       0    &         2       &   100& 1.5--4&   150 $\pm$ 20    \\
                                  
Lupus III     &     340, 9     &      15     &      10   &          5      &   33  &1.5--4 & 200 $\pm$ 20\\
                                  
Lupus IV      &    336, 8    &       5     &    2     &         3      &   60   & 1.5--4&150 $\pm$ 20\\
                                  
Lupus V       &    342, 9      &      16       &    0     &         16        &  100 & 10 & 150 $\pm$ 20  \\ 
                                  
Lupus VI      &    342, 6      &     15      &      0     &          15     &   100 & 10  &150 $\pm$ 20\\
                                  
Cra            &    0,-19       &    5          &    4          &      1      &  20  & 1  &150 $\pm$ 20\\
                                  
Scp            &    250,18       &    2         &    1           &     1     &   50   & 5  &130 $\pm$ 20\\
                                  
Oph (Paper I)           &  353,18       &    34          &  26        &     8       &  24    &  2 &150 $\pm$ 20\\
\enddata
\end{deluxetable}

\begin{table*}
\begin{center}
\caption{\label{giants-all}
{\small{\bf Distribution of YSOc, transition disks (TD) and AGBs candidates organized by \emph{c2d} and \emph{Gould Belt} Legacy Projects' clouds}}}
\begin{tabular}{c|c|c|c|c}                           
\hline\hline
\small
\small {\bf Region} &  YSOc & AGB &  TD   & AGB \\
                     &            & candidates          & candidates     &
candidates in TD region  \\
                    &            & whole sample         &      &   in TD region   \\
                    &\#      & \%          &\#            & \%                         \\
\hline
\multicolumn{5}{c}{\emph {c2d} Legacy Project's clouds}\\
\hline
{\bf CHA II}  & 29          &  10.4     & 7  & 28.6 \\
{\bf LUP I}   & 20          &  15       & 8  & 25\\
{\bf LUP III} & 79          &    18.9   & 18 &11.1   \\
{\bf LUP IV}  & 12          &  25       & 5   &  20  \\
{\bf OPH}     & 297         &  7.7      &  52&     15.38  \\  
{\bf PER}     & 387         &  2.6      &  56   &   10.7 \\         
{\bf SER}     &  262         &  6.5     &  60   &   10    \\
\hline
\multicolumn{5}{c}{\emph{Gould Belt} Legacy Project's clouds}\\
\hline
{\bf AURIGA}  & 174        &  1.7          &  28 & 7.1   \\
{\bf CrA}     & 45          & 4.4          &  7 & 14.2    \\
{\bf IC5146}  & 163        & 2.4           &  24 &  4.1 \\
{\bf LUP V}   & 44         & 47.7          & 22   &   36.3   \\ 
{\bf LUP VI} & 46         & 67.3          &  21   &   57.1        \\
{\bf SERP-AQUILA} & 1442    & 28.6         &641 & 32.6\\
{\bf CHAM I} &  93           & 1           &17 &  5.9\\ 
{\bf CHAM III} & 4           & 75           &1  & 100  \\
{\bf MUSCA} &  13             &84.6    & 5  &       80.0 \\  
{\bf CEPH} &  119               &2.5 & 19  &  10.5\\
{\bf SCO} &   9                  &11 & 4 & 25 \\
\hline
\end{tabular}
\end{center}
\end{table*}

\begin{landscape}
\begin{deluxetable}{rrcrrrrrrrrrrrc}
%\rotate
\tablewidth{0pt}
\tabletypesize{\tiny}
\tablecaption{\bf{Transition Disk Sample}}
\tablehead{\colhead{\#}&\colhead{\emph{Spitzer} ID}&\colhead{Alter. Name}&\colhead{R$_1$}&\colhead{R$_2$}&\colhead{J\tablenotemark{a}}&\colhead{H}&\colhead{K$_{S}$}&\colhead{F$_{3.6}$\tablenotemark{a}}&\colhead{F$_{4.5}$}&\colhead{F$_{5.8}$}&\colhead{F$_{8.0}$}&\colhead{F$_{24}$}&\colhead{F$_{70}$\tablenotemark{b}  }&\colhead{Region}   \\
\colhead{}&\colhead{}&\colhead{}&\colhead{(mag)}&\colhead{(mag)}&\colhead{(mJy)}&\colhead{(mJy)}&\colhead{(mJy)}&\colhead{(mJy)}&\colhead{(mJy)}&\colhead{(mJy)}&\colhead{(mJy)}&\colhead{(mJy)}& \colhead{(mJy)} & }
\startdata  
1 &  SSTc2d\_J160026.1-415356 &      .....    & 15.62& 15.54 & 2.97e+01  &3.70e+01  &3.25e+01  &2.15e+01 & 1.68e+01 & 1.42e+01 & 1.63e+01 & 2.40e+01 &$<$ 50 &   Lup  IV   \\
2 &  SSTc2d\_J160044.5-415531 &      V*MYLup  & 11.22& 11.06 & 2.63e+02  &3.44e+02  &3.05e+02  &1.77e+02 & 1.41e+02 & 1.40e+02 & 2.13e+02 & 5.90e+02 &1.05e+03&  Lup  IV \\
3 &  SSTc2d\_J160711.6-390348 &      SZ91     & 13.61& 13.89 & 6.03e+01  &9.13e+01  &7.67e+01  &3.86e+01 & 2.47e+01 & 1.72e+01 & 1.09e+01 & 9.72e+00 &5.02e+02&  Lup  III  \\
4 &  SSTc2d\_J160752.3-385806 &      SZ95     & 13.66& 14.02 & 6.28e+01  &7.89e+01  &6.61e+01  &4.21e+01 & 3.18e+01 & 2.73e+01 & 2.96e+01 & 3.00e+01 &$<$ 50 &     Lup  III \\
5 &  SSTc2d\_J160812.6-390834 &      SZ96     & 12.98& 13.66 & 1.42e+02  &1.87e+02  &1.74e+02  &1.68e+02 & 1.13e+02 & 1.38e+02 & 1.73e+02 & 2.41e+02 &1.54e+02&  Lup  III   \\
6 &  SSTc2d\_J160828.4-390532 &      SZ101   &  13.52& 13.53&   1.10e+02&  1.32e+02& 1.17e+02 & 7.98e+01&  5.55e+01&  4.16e+01&  3.29e+01&  2.41e+01& $<$ 50&     Lup  III \\
7 & SSTc2d\_J160831.5-384729 &      Lup\,338 & 12.70& 13.03 &  2.15e+02 &
2.75e+02 & 2.37e+02 & 1.49e+02&  9.25e+01&  6.98e+01&  5.16e+01&  2.85e+01&
$<$ 50&   Lup  III    \\
8 & SSTc2d\_J160841.8-390137 &      SZ107    & 15.29& 15.47 &  5.05e+01 & 5.76e+01 & 5.01e+01 & 2.65e+01&  2.02e+01&  1.41e+01&  9.24e+00&  1.07e+01& $<$ 50&     Lup  III  \\
 9 & SSTc2d\_J160855.5-390234 &      SZ112    & 14.57& 14.68 &  6.32e+01 &
 7.87e+01 & 6.90e+01 & 4.87e+01&  3.80e+01&  3.04e+01&  2.48e+01&  1.24e+02&
 1.20e+02&Lup  III    \\
 10 & SSTc2d\_J160901.4-392512 &      ....     & 14.83& 14.89 &  3.62e+01 & 5.45e+01 & 5.09e+01 & 4.47e+01&  3.40e+01&  3.06e+01&  2.58e+01&  4.22e+01 & 1.14e+02& Lup  III    \\
 11 & SSTc2d\_J160954.0-392328 &      Lup\,359 & 12.96& 13.30 &  1.59e+02 & 2.13e+02 & 1.94e+02 & 1.35e+02&  1.01e+02&  8.44e+01&  7.93e+01&  9.65e+01& $<$ 50&     Lup  III  \\
 12 & SSTc2d\_J161029.6-392215 &      ...      & 15.69& 15.79 &  2.66e+01 & 3.18e+01 & 2.88e+01 & 1.91e+01&  1.42e+01&  1.15e+01&  1.09e+01&  3.37e+01& 1.10e+02& Lup  III    \\
 13 & SSTc2d\_J162209.6-195301 &   ...            &14.48 &14.27  & 1.44e+02  &2.08e+02  &1.84e+02  &1.09e+02 & 7.15e+01 & 5.05e+01 & 3.27e+01 & 1.59e+01 & $<$ 50&     Scp \\
 14 & SSTGBS\_J190029.1-365604 &       CrAPMS8 &13.80 &13.78  & 7.84e+01
 &1.06e+02  &9.89e+01  &5.08e+01 & 3.60e+01 & 2.62e+01 & 1.83e+01 & 3.59e+01 &
 $<$ 100 &     Cra\\
 15 & SSTGBS\_J190058.1-364505 &  CrA-9            &13.49 &13.57  & 1.12e+02
 &1.61e+02  &1.40e+02  &5.81e+01 & 4.38e+01 & 3.18e+01 & 2.48e+01 & 1.78e+02 &
 $<$ 100 &     Cra\\
 16 & SSTGBS\_J190129.0-370148 &    G-94           &15.53 &14.95  & 3.53e+01  &4.25e+01  &3.62e+01  &1.95e+01 & 1.38e+01 & 9.67e+00 & 6.52e+00 & 2.92e+00 & $<$ 50&     Cra \\ 
 17 & SSTGBS\_J190311.8-370902 &   CrA-35         &17.20 &16.82  & 2.46e+01  &3.24e+01  &3.09e+01  &2.12e+01 & 1.58e+01 & 1.21e+01 & 1.06e+01 & 1.18e+01 & $<$ 50&     Cra\\
\enddata  
\tablenotetext{a}{All the 2MASS, IRAC, and 24 $\mu$m detections are $\ge$ 7$\sigma$ (i.e.,the photometric uncertainties are $\lesssim$~15$\%$)}
\tablenotetext{b}{$\ge$ 5$\sigma$ detections from the \emph{c2d} and \emph{Gould-Belt} catalogs or 5$\sigma$ upper limits as  described in \S~\ref{sed_mor}.}   \label{t:mags}                  
\end{deluxetable}
\end{landscape}   
 
\begin{deluxetable}{rcccrcrrrcc}
% \rotate
\tablewidth{0pt}
% \tablewidth{\textwidth}
\tabletypesize{\scriptsize}
\setlength{\tabcolsep}{0.8ex}
 \tablecaption{{\bf Observed Properties}}
 \tablehead{\colhead{\#}&\colhead{Ra (J2000)}&\colhead{Dec (J2000)}&\colhead{Tel. SpT}&\colhead{SpT.}&\colhead{References\tablenotemark{a}}&\colhead{Li I\tablenotemark{b}}&\colhead{H$\alpha$\tablenotemark{b,c}}&\colhead{Flux$_{mm}$\tablenotemark{d}}&\colhead{$\sigma$Flux$_{mm}$}&\colhead{Separ\tablenotemark{e}}\\
\colhead{}&\colhead{(deg)}&\colhead{(deg)}&\colhead{}&\colhead{}&\colhead{}&\colhead{(\AA)}&\colhead{(km s$^{-1}$)}&\colhead{(mJy)}&\colhead{(mJy)}&\colhead{(arcsec)}}
\startdata 
 1 &  240.10887       & -41.89877 & Clay      &  M5.25,M1$^f$                  & 1,2 &      0.47 &   162    & $<$  21   &  7    &2.8  \\
 2 &  240.18554       & -41.92534 &  Du\,Pont &    K0                    &   7 &   0.44    &   532    &  100  &  5    &        \\
 3 &  241.79833       & -39.06326 &  Du\,Pont &  M1.5,M0.5                 & 1,6   &     0.41  &      283 & 34.5 &  2.9  &5      \\
 4 &  241.96800       & -38.96840 &  Du\,Pont & M3.25,M1.5                   & 1,6   &   0.46    &    321   & $<$  9.9    & 3.3   &3     \\
 5 &  242.05258       & -39.14264 &  Du\,Pont &   M2,M1.5                & 1,6   &   0.5     &   233    &  $<$ 21    &  7    &      \\
 6 &  242.11837       & -39.09229&   Du\,Pont&   M5,M4                 &  1,6  &   0.28    &  343     &   $<$ 10.8  &  3.6  & 0.7   \\
 7 & 242.13146       & -38.79148 &  Du\,Pont &   M2.25,M2                  & 1,4   &   0.25    &  382     &    6.7  & 2.2   & 0.4   \\
 8 & 242.17413       & -39.02695 &  Du\,Pont &   M5.75,M5.5                & 1,6   &  \nodata  & 200      &  9.7 & 2.5    &    \\
 9 & 242.23133       & -39.04276 &  Du\,Pont &   M6,M6                  & 1,6   &  \nodata  &    189   &    $<$ 21  &   7   &2.8    \\
 10 & 242.25583       & -39.41997 &  Clay     & M4,M4                    & 1,2 &     0.45  &    369   &  31.4&   3.4 &        \\
 11 & 242.47496       & -39.39109 &Du\,Pont   &  M2.75,M1.5                 & 1,4   &    0.40   &    336   &  16.7&   3.3 & 1.15    \\
 12 & 242.62321       & -39.37076 &  Clay     &  M4.5,M4                  & 1,2&     0.52  &     180  &  23.2     &  4.7  &         \\
 13 & 245.54000       & -19.88357  &  Clay     &    M3.7                      &1   &  0.55     & 132      & $<$ 10.8     & 3.6   &1.8,3      \\
 14 & 285.12113       & -36.93437  & Du\,Pont  &    M4,M3                  &1,5   &0.30       & 93       & $<$ 10.5     &  3.5 &0.132   \\
 15 & 285.24187       & -36.75139  & Clay      &    M0.75                     & 1  &      0.48 &     440  &    $<$ 21  &   7   &       \\
 16 & 285.37088       & -37.03011  &  Du\,Pont &    M3.75,M3.5                &1,3   &   \nodata &    83    & $<$ 21      &  7    &0.5   \\
 17 & 285.79929       & -37.15055  &  Clay     &    M5.0                     &1       & 0.51      &  205   &      $<$ 21   &     7 &0.5   
\enddata                                        
\tablenotetext{a}{References. (1) Spectral type from this work; (2) \citet{merinetal08-1}; (3) \citet{Sicilia-Aguilaretal08-1}; (4) \citet{krautteretal97-1}; (5) \citet{walteretal97-1} ;(6) \citet{hughesetal94-1}; (7) \citet{hughesetal93-1}}
\tablenotetext{b}{``\nodata" implies that the signal to noise in this region of the spectrum is too low to measure the width or establish the presence of the line}.
\tablenotetext{c}{"-1" implies that H$\alpha$ is seen in absorption.}
\tablenotetext{d}{"$<$" implies upper limits value (3 $\sigma$)}
\tablenotetext{e}{Sources \# 1, 3, 4, and 9 have been identified as 
binaries by WFI observations \citep{merinetal08-1}. The binary nature of 
source \# 14 was discovered using speckle interferometry at the NTT (La Silla) 
revealing a projected separation of 
0.132$''$ $\pm$ 0.009$''$ \citep{kohleretal08-1}.
Sources \# 11, 13 are triple systems with tight binary 
components consistent with 
two equally bright objects and a projected separation 
of $\sim$0.05$''$.}
\tablenotetext{f}{The spectral type derived by us is significantly
    later than the previously obtained value. We consider our 
estimate of M5.25 based on a high quality optical spectrum 
to be more reliable than the rather rough guess 
of \citet{merinetal08-1} derived from optical and near-IR photometry only.}
 \label{t:obs}
\end{deluxetable}

\begin{deluxetable}{lrrrrrcrc}
\tabletypesize{\scriptsize}
\tablewidth{0pt}
\tablecaption{{\bf Derived Properties}}
\tablehead{\colhead{\#}&\colhead{LOG(Acc. rate)}&\colhead{Mass
    Disk}&\colhead{r$_{proj.}$\tablenotemark{a}}&\colhead{$\lambda_{tun-off}$}&\colhead{$\alpha_{excess}$}&\colhead{LOG(L$_{D}$/L$_{*}$)}&\colhead{L$_{*}$}&\colhead{Classification}\\
\colhead{}&\colhead{(M$_{\odot}$/yr)}&\colhead{(M$_{JUP}$)}&\colhead{(AU)}&\colhead{$\mu$m}&\colhead{}&\colhead{}&\colhead{\Lsun}&\colhead{Cloud partnership}}
\startdata
 1   &    $<$ -11     & $<$  1.9  &      420           &    4.50 &    -0.80  &-2.1 &  0.06  &   photo.  disk, Lup IV\\
 2   &    -7.7        &  9.1      &                    &    4.50 &
-0.17     &-2.4 & 2.37    & giant planet,  Lup IV \\ 
 3   &     -10.1      & 5.6       &      1000          &    8.00 &
-2.18  &  -2.6 & 0.39   & giant planet,  Lup III  \\ 
 4    &    -9.7       &  $<$ 1.6  &      600          &    4.50 &
-1.05  & -2.2& 0.30    & grain growth$^b$,  Lup III   \\ 
 5    &     -10.6     &  $<$ 3.4  &                   &    2.20  &
-0.93 & -1.4   &  0.86 & grain growth$^b$,  Lup III   \\ 
 6  &      -9.5       &  $<$ 1.8  &      140       &    5.80  &    -1.42
& -2.6   &  0.43 & grain growth,  Lup III   \\ 
 7    &     -9.1       & 1.1       &          80      &    8.00 &
-1.55  &  -2.9 & 1.27   & grain growth,  Lup III   \\ 
 8   &      -11       &  1.6      &     76             &       5.8&
-0.86 &  -2.8   & 0.17 & grain growth$^b$,  Lup III   \\ 
 9   &     $<$ -11    &   $<$ 3.4 &     560            &    4.50 &
-0.31  & -1.9    &  0.22&  photo. disk,  Lup III \\
 10   &     -9.3       & 5.1       &                    &    2.20  &
-1.19  &   -1.5  & 0.27 & grain growth$^b$,  Lup III   \\ 
 11   &     -9.6       &  2.7      &      230\nodata$^a$     &    4.50
&    -1.05  &  -2.3  & 0.96  &circumbinary/gr-grow$^b$, Lup III\\ 
 12   &     -11        &  3.8        &                    &    5.80  &
-0.28  &   -2.1   &  0.11 &  grain growth$^b$,  Lup III \\
 13   &     $<$ -11    &  $<$ 0.8  &     234,390$^a$      &    8.00 &
-1.67  &  -3.7    & 0.48 &  circumbinary/debris, Scp     \\
 14   &     $<$ -11    &  $<$ 1    &      20            &    5.80 &
-0.42  & -2.7      &  0.14 & circumbinary/ photo. disk, CrA    \\
 15   &      -8.6      &  $<$ 2  &                    &    8.00 &
0.76  & -2.4   &   0.46 & giant planet, CrA \\
 16  &     $<$ -11     & $<$ 2   &      75            &    8.00 &
-1.74  &  -3.2   &  0.07 & debris disk, CrA  \\
 17   &      -11       & $<$ 2   &      75            &    5.80  &
-1.06  &  -2.3   &  0.06 & grain growth, CrA   \\
\enddata
\tablenotetext{a}{Objects \# 11 and 13 are triple systems. The
    distances given correspond to the closest and widest components. For the
    case of \# 11, the tightest components can not be resolved.}
\tablenotetext{b}{Uncertain classification due to relatively 
weak evidence for accretion (target \# 8 and 12) or SEDs similar to those of
classical T\,Tauri stars of spectral type M (target \# 4, 5, 10 and 11). }
\label{t:derived}
\end{deluxetable}

\begin{table*}
\begin{center}
\small \caption{\label{t:fractions}
{\bf Fractional distribution of transition disk organized by clouds}}
\vspace{0.2cm}
\begin{tabular}{c|c|c|c}
\hline\hline
{\bf Disk candidates} &{\bf  Lupus~III,~IV} &{\bf Cra}& {\bf Oph$^\dag$} \\
\hline
{\bf debris}          &  --              & 1 (25\%) & 4 (15\%) \\
{\bf photoevaporated}  & 2  (17\%)        & 1 (25\%) &   5 (19\%)\\
{\bf grain growth}     &  8    (66\%)         &  1 (25\%)  & 13 (50\%)\\
{\bf hosting giant planets}  &   2 (17\%)     &   1 (25\%)  &  4 (15\%)\\
{\bf circumbinary}       &     1  (8\%)         & 1 (25\%)    & 4 (15\%)         \\
\hline
{\bf Total}          &  12   & 4 &26\\
\hline
{\bf Age [Myr]}            & $\sim$ 1.5--4$^a$     & $\sim$ 1$^b$       & $\sim$  2$^c$  \\
\hline
\end{tabular}
\end{center}
$^\dag$ From Paper I\\
References: $^a$  \citet{comeron08-1};  $^b$ \citet{wilkingetal05-1}; $^c$ \citet{Sicilia-Aguilaretal08-1}
\end{table*}

\end{document}